\documentclass[12pt, preprint]{aastex}

\begin{document}

\newcommand{\lya}{{\rm Ly}$\alpha$}
\newcommand{\Lya}{{\rm Ly}$\alpha$}
\newcommand{\lyb}{{\rm Ly}$\beta$}
\newcommand{\Lyb}{{\rm Ly}$\beta$}

\newcommand{\NHI}{\mbox{N(\HI )}}
\newcommand{\NDI}{\mbox{N(\HI )}}
\newcommand{\nsubhavg}{\mbox{$\langle N_H \rangle$}}
\newcommand{\nhi}{\mbox{N(\HI )}}
\newcommand{\ndi}{\mbox{N(\HI )}}
\newcommand{\di}{\mbox{\ion{D}{1}}}
\newcommand{\hi}{\mbox{\ion{H}{1}}}
\newcommand{\HI}{\mbox{\ion{H}{1}}}
\newcommand{\DI}{\mbox{\ion{D}{1}}}
\newcommand{\CI}{\mbox{\ion{C}{1}}}
\newcommand{\CIstar}{\mbox{\ion{C}{1}$^*$}}
\newcommand{\CII}{\mbox{\ion{C}{2}}}
\newcommand{\CIIstar}{\mbox{\ion{C}{2}$^*$}}
\newcommand{\CIII}{\mbox{\ion{C}{3}}}
\newcommand{\CIV}{\mbox{\ion{C}{4}}}
\newcommand{\NI}{\mbox{\ion{N}{1}}}
\newcommand{\NII}{\mbox{\ion{N}{2}}}
\newcommand{\NV}{\mbox{\ion{N}{5}}}
\newcommand{\OI}{\mbox{\ion{O}{1}}}
\newcommand{\oi}{\mbox{\ion{O}{1}}}
\newcommand{\OIV}{\mbox{\ion{O}{4}}}
\newcommand{\OV}{\mbox{\ion{O}{5}}}
\newcommand{\OVI}{\mbox{\ion{O}{6}}}
\newcommand{\NaI}{\mbox{\ion{Na}{1}}}
\newcommand{\SiII}{\mbox{\ion{Si}{2}}}
\newcommand{\SiIIstar}{\mbox{\ion{Si}{2}}$^*$}
\newcommand{\SiIII}{\mbox{\ion{Si}{3}}}
\newcommand{\SiIV}{\mbox{\ion{Si}{4}}}
\newcommand{\AlII}{\mbox{\ion{Al}{2}}}
\newcommand{\FeI}{\mbox{\ion{Fe}{1}}}
\newcommand{\FeII}{\mbox{\ion{Fe}{2}}}
\newcommand{\FeIII}{\mbox{\ion{Fe}{3}}}
\newcommand{\PII}{\mbox{\ion{P}{2}}}
\newcommand{\ArI}{\mbox{\ion{Ar}{1}}}
\newcommand{\HH}{\mbox{H$_2$}}
\newcommand{\Ht}{\mbox{H$_T$}}
\newcommand{\DHrat}{\mbox{\rm D/H}}
\newcommand{\DHrattot}{\mbox{\rm [\DI\ + HD]/[\HI\ + 2\HH ]}}
\newcommand{\DO}{\mbox{\rm D/O}}
\newcommand{\DOtot}{\mbox{\rm [\DI\ + HD]/\OI }}
\newcommand{\DN}{\mbox{\rm D/N}}
\newcommand{\DNtot}{\mbox{\rm [\DI\ + HD]/\NI }}
\newcommand{\OH}{\mbox{\rm O/H}}
\newcommand{\OHtot}{\mbox{\rm O/[H+2\HH ]}}
\newcommand{\NH}{\mbox{\rm N/H}}
\newcommand{\NHtot}{\mbox{\rm N/[H+2\HH ]}}
\newcommand{\NO}{\mbox{\rm N/O}}
\newcommand{\hmol}{\mbox{H$_2$}}
\newcommand{\pgoo}{\mbox{PG~0038$+$199}}
\newcommand{\dhprim}{\mbox{{\rm D/H}}}
\newcommand{\dhld}{\mbox{\rm D/H}}
\newcommand{\dhldg}{\mbox{\rm D/H}}
\newcommand{\vpfit}{\mbox{\sc vpfit}}
\newcommand{\owens}{\mbox{\sc Owens}}
\newcommand{\Owens}{\mbox{\sc Owens}}
\newcommand{\chisq}{\mbox{$\chi^2$}}
\newcommand{\fuse}{\mbox{\it FUSE}}
\newcommand{\FUSE}{\mbox{\it FUSE}}
\newcommand{\hst}{\mbox{\it HST}}
\newcommand{\HST}{\mbox{\it HST}}
\newcommand{\iue}{\mbox{\it IUE}}
\newcommand{\IUE}{\mbox{\it IUE}}
\newcommand{\ecs}{\mbox{$\rm erg~cm^{-2} sec^{-1}$}}

\newcommand{\simlt}{{\la}}
\newcommand{\simgt}{{\ga}}
\newcommand{\kms}{km~s$^{-1}$}
\newcommand{\cmsq}{cm$^{-2}$}

\slugcomment{accepted by ApJ, 15 Jan 2005}
\pagestyle{myheadings}

\title{The D/H Ratio toward PG~0038+199$^{1,2}$}

\footnotetext[1]{This paper is dedicated in memory of Ervin J. Williger, father of the first author,
who passed away on 2003 September 13.  His enthusiastic support and encouragement were essential to its successful
completion.}
\footnotetext[2]{Based on data from the {\it Far Ultraviolet Spectroscopic Explorer} and the W. M. Keck Observatory}

\author{Gerard M. Williger, Cristina Oliveira}
\affil{Dept. of Physics \& Astronomy, Johns Hopkins Univ., Baltimore MD 21218}

\author{Guillaume H\'ebrard }
\affil{Institut d'Astrophysique de Paris, 98 bis, Blvd Arago, 75014 Paris, France and 
\\Dept. of Physics \& Astronomy, Johns Hopkins Univ., Baltimore MD 21218}

\author{Jean Dupuis}
\affil{Dept. of Physics \& Astronomy, Johns Hopkins Univ., Baltimore MD 21218}

\author{Stefan Dreizler\altaffilmark{3} }
\affil{Institut f\"ur Astronomy und Astrophysik, Waldh\"auserstr. 64, 72076 T\"ubingen, Germany}
\altaffiltext{3}{present address: Uni-Sternwarte, Geismarlandstr. 11, 37083, G\"ottingen, Germany}

\author{H. Warren Moos}
\affil{Dept. of Physics \& Astronomy, Johns Hopkins Univ., Baltimore MD 21218}

\begin{abstract}

We determine the D/H ratio 
in the interstellar medium toward the
DO white dwarf \pgoo\ using spectra from the {\it Far Ultraviolet
Spectroscopic Explorer} (\FUSE ), with ground-based support from Keck HIRES.  
We 
employ curve of growth, apparent optical depth and profile fitting techniques to
measure column densities and limits of many other
species (\HH , \NaI , \CI , \CII , \CIII , \NI , \NII , \OI , \SiII , 
\PII , \ion{S}{3}, \ArI\ and \FeII ) which allow us
to determine related ratios such as D/O, D/N and the \HH\ fraction.
Our efforts are concentrated on measuring gas-phase D/H, which is key to
understanding Galactic chemical evolution and comparing it to predictions from
Big Bang nucleosynthesis.  
We find column densities $\log N(\HI ) = 20.41\pm 0.08$,
$\log N(\DI ) = 15.75\pm 0.08$ and $\log N(\HH ) = 19.33\pm 0.04$,
yielding a molecular hydrogen fraction of $0.14\pm 0.02$ ($2\sigma$ errors), with
an excitation temperature of $143\pm 5$~K.  The high \HI\ column density implies that
\pgoo\ lies outside of the Local Bubble; we estimate its distance 
to be $297^{+164}_{-104}$~pc ($1\sigma$).
\DHrattot\ toward \pgoo\ is
$1.91^{+0.52}_{-0.42}\times 10^{-5}$ ($2\sigma$).  
There is no evidence of component structure on the scale of $\Delta v > 8$~\kms\
based on \NaI , but there is marginal evidence for structure on smaller scales.
The D/H value is high compared to the majority of recent D/H measurements, but
consistent with the values for two other measurements at similar distances. 
D/O is in agreement with other distant measurements.
The scatter in D/H values 
beyond $\sim 100$~pc remains a challenge for
Galactic chemical evolution.
\end{abstract}


\section{Introduction}

The ratio of D/H in the interstellar medium (ISM) offers a means to  measure
the effects of supernovae, stellar winds, infalling halo gas and ISM 
mixing on the
chemical evolution of the Galaxy.  The primordial D/H ratio is a firm
prediction of the cosmic baryon density from Big Bang nucleosynthesis
\citep{Schramm98,Burles00,Burles01}.   Measuring the D/H ratio in environments as close to
primordial as possible in practice means making determinations in
low metallicity, high redshift QSO absorbers. \citet{Kirkman03}, building on
published work \citep{Omeara01,Pettini01,Levshakov02} determined primordial
$\dhprim =(2.78^{+0.44}_{-0.38})\times 10^{-5}$  ($1\sigma$)\footnote{We will often
distinguish between $1\sigma$ and $2\sigma$ errors in this work, given the
variety of errors cited in the literature and some conventions observed in
previous papers dealing with \DHrat .  Unless otherwise stated,
errors are $1\sigma$.} which is consistent with WMAP results
and Big Bang nucleosynthesis \citep{Romano03,Cyburt03}. Closer to the Milky
Way but presumably outside of the Galactic disk, the D/H ratio from the high
velocity cloud Complex C is ($2.2\pm 1.1)\times 10^{-5}$ ($2\sigma$)
\citep{Sembach04}, slightly lower than but still consistent with the
primordial value.  

D/H ratios should be lower than the primordial value
in regions where star formation has taken place,  and thus, destruction of \DI\  
occurred in stellar interiors (astration).  However, fluctuations in
the D/H ratio may occur due to a number of reasons.  Typical processes
that could produce such \DHrat\ variations would include inhomogeneous
ISM
mixing by supernovae, stellar winds and the infall of relatively unastrated matter
into the Galaxy.   In addition, recent studies have been made of differential depletion 
of HD onto polycyclic aromatic
hydrocarbons (PAHs) or dust \citep{Draine04,Lipshtat04,Peeters04}, modulated
over scales of hundreds of pc and several Myr \citep{deAvillez02}.

To probe the
pattern of \DI\ processing, measurements are being
made of the gas-phase \DHrat\ ratio at a variety of distances.
On the nearest scale,
\citet{Linsky98} found $\DHrat =(1.5\pm0.1)\times10^{-5}$ from 12
sight lines within the Local Interstellar Cloud (LIC). Further out in the
Local Bubble, which is an irregular cavity filled with hot, low density, ionized gas
with a radius $\sim 65-250$~pc \citep{Sfeir99}, four sight lines are consistent
with the LIC value \citep[][and references therein]{Oliveira03,Moos02}.
\citet{Wood04} determined $\DHrat=(1.56\pm 0.04)\times10^{-5}$ ($1\sigma$ standard
deviation of the mean) for 22 sight lines in the Local Bubble compiled from the literature.
\citet{Hebrard03} used
14 sight lines to find a value 
of $\DHrat = (1.32\pm 0.08)\times 10^{-5}$ 
from \DO\  
\citep[building on the O abundance studies of][]{Meyer98}, which is
marginally 
in agreement  with the LIC value.\footnote{A common implicit assumption is that
the local disk value for \DHrat\ is approximately the same as the gas phase
value. This misses any deuterium which is locked into grains or HD
molecules, which may or may not be important.  In one example, 
\citet{Jenkins99} do not find HD to be a significant
reservoir of D toward $\delta$~Ori, which has a low \DHrat\ ($0.74^{+0.19}_{-0.13}$,
90\% confidence) and no HD lines in their IMAPS data 
\citep[$\log N({\mbox{\rm HD}})<12.8$; ][]{Spitzer74}.
\citet{Lacour04} also find that HD is not  a significant reservoir of
D within several kpc for a sample of seven stars. The issue is under current
debate in the literature \citep{Ferlet00,Lipshtat04}.}

At distances between the Local Bubble and Complex C
($0.1\simlt d \simlt 3$~kpc),  however, \dhld\ does show a dispersion of values. 
The lines of sight toward 
$\lambda$~Sco, $\delta$~Ori, $\epsilon$~Ori, LSS~1274, JL~9, HD~195965 and
HD~191877 \citep{York83,Jenkins99,York76,Laurent79,Wood04,Hoopes03} were
measured with {\it Copernicus}, the Interstellar Medium Absorption Profile
Spectrograph (IMAPS) and \fuse\ to have low \DHrat\ values of 
$0.65<(\dhld )/10^{-5} <1.00$, with
errors of $\simlt 0.3$ in D/H for each measurement.  In sharp contrast, 
Feige~110 \citep{Friedman02} and $\gamma^2$~Vel \citep{Sonneborn00} were measured
with IMAPS and \FUSE\ to have high values of 
$\DHrat=(2.18^{+0.36}_{-0.31})\times 10^{-5}$
(90\% confidence limits) and $\DHrat=(2.14\pm 0.82)\times 10^{-5}$
($2\sigma$), respectively.
Although the  number of \dhld\ measurements outside the Local Bubble is not high enough
to characterize the nature of the \dhld\ dispersion,  these
two anomalously high values of \dhld\ could indicate a
significant inhomogeneity in the ISM.

In principle, such inhomogeneities should be averaged out if one measures \dhld\ over
long enough sight lines.  However, confusion between velocity components  at
differing distances makes such measurements increasingly difficult for high \hi\
column densities.   \citet{Hebrard03} used D/O and D/N measurements as proxies for
\DHrat\ toward stars at distances $d\simgt 100$~pc, and concluded that \dhld\ is
generally lower   outside of the Local Bubble.  D/O and D/N measurements avoid the
need to measure the \hi\ column density, which can be difficult depending on whether
the Lyman series falls largely on the flat part of the curve of growth and/or whether
data for \Lya\ are available from \HST\ or \IUE .  \DI\ and \OI\ also have column
densities that only differ by $\sim 2$ orders of magnitude, which reduces systematic
errors compared to \DHrat , 
where the difference
in column densities is 
$\sim 5$ orders of magnitude. 
The D/O ratio has its own complexities, though, due to a dearth of unsaturated \oi\
lines for high column densities, and the possibility of uncertain \OI\ oscillator
strengths, whereas at the highest column densities, $N(\HI )$ becomes easier to
measure with the onset of Ly$\alpha$ and Ly$\beta$  damping wings \citep{Wood04}.

In light of the \dhld\ and D/O measurements, it is therefore crucial to
characterize the variations in \DI\ abundance outside of the Local Bubble,
both as a probe of astration, and as a means to constrain processes such
as ISM mixing, infall of near 
primordial gas from the intergalactic medium and satellite galaxies
and deuterium depletion.

White dwarfs produce simple continua that act as ideal background sources for studying
ISM absorption lines, and thus provide useful sight lines out to a few hundred
pc. In this paper, we describe \DHrat\ measurements using \FUSE\ observations
of the hot DO white dwarf \pgoo . We present the target data in
\S~\ref{sec-pg0038}.  Details of observations and data reduction, including
profile fitting (PF), curve-of-growth (COG) and apparent optical depth (AOD)
techniques are in \S~\ref{sec-observations}.   We describe our analysis in
\S~\ref{sec-analysis}, discuss the results in \S~\ref{sec-discussion} and
summarize our conclusions in \S~\ref{sec-conclusions}.

\section{PG 0038+199}
\label{sec-pg0038}

The DO white dwarf \pgoo\ was discovered via UV excess \citep{Green86}.
\citet{Wesemael93} classified it as a hot DO star, and \citet{Dreizler96}
performed a model atmosphere analysis. 
\citet{Dreizler97} found $\log {\rm (H/He)}=-0.7$
from Keck HIRES data, though the \HI\ emission line is very weak.
\citet{Kohoutek97} listed the object as a possible post-planetary nebula.
However, \citet{Werner97} searched for an old planetary nebula in the region
by direct H$\alpha$ imaging, and found no evidence for one.  A summary of stellar
parameters is in Table~\ref{tab-stellarprops}.

The distance to \pgoo\ is estimated 
using the photometric parallax technique. 
Using a magnitude $V=14.544$ 
and color excess E(B-V)=0.037 (Table~\ref{tab-stellarprops}) and
synthetic spectra computed in the optical,
we calculate
a distance of $297^{+164}_{-104}$~pc (with the errors
dominated by the uncertainty in stellar gravity; 
see \S~\ref{sec-stellarmodelpg0038} for model details).

\pgoo\ (Galactic $\ell = 119.79^\circ$, $b=-42.66^\circ$) is
outside of the Local Bubble, and is bright in the far UV based on IUE spectra.
It is therefore
a good candidate to help characterize the dispersion in \DHrat\ at large distances.
Low dispersion \IUE\ spectra of \pgoo\ exist, but are not suitable for measuring the \HI\
column density from the \Lya\ line.  There are no other 
high resolution UV spectra of \pgoo\ in the Multimission Archive at 
the Space Telescope Science Institute except for
the \FUSE\  observations presented here.

\section{Observations and Data Reduction}
\label{sec-observations}

\subsection{Keck observations and reduction}

High resolution optical spectra were obtained with the HIRES echelle 
spectrograph \citep{Vogt94} on the Keck I telescope on 1996 July 24,
using the red cross disperser to cover the wavelength region between 
4260 \AA\ and 6700 \AA\ at a resolution 8.3 \kms . The exposure time was 2000~sec.
The standard data reduction as described by \citet{Zuckerman98}
resulted in spectral orders that have a somewhat wavy continuum. 
To remove the echelle ripple, we used the spectrum of BD+33$^\circ $2642 
which was observed in the same night. Data were corrected to the vacuum heliocentric frame.
\NaI\ was used as an absolute velocity reference for the neutral
gas in the \FUSE\ data.

\subsection{\boldmath{$FUSE$} observations}
\pgoo\ was observed by \FUSE\
for 13100~sec on 2000 July 27 (data set P1040201)
through the large aperture (LWRS) in time-tag mode.  
For a description of the \FUSE\ instrument see \citet{Moos00} and \citet{Sahnow00}.
Focal plane splits were not performed, because it was early in the \FUSE\
mission and they were not routinely done at the time for exposures of
that length.
It was also observed for 6600~sec for 
program M114, a calibration program for periodic monitoring of channel
co-alignment, but those observations involved motions
in the direction of dispersion, and therefore were not deemed scientifically
useful for this work.

Because the observations were done with the LWRS aperture, airglow
is significant in the troughs of \hi\ Ly$\beta$ and Ly$\gamma$, with 
telluric \oi ~$\lambda 1027.5$~\AA\ also affecting the Ly$\beta$ damping wing.
We excluded those regions from our analysis.

\subsection{\boldmath{$FUSE$} data reduction}

\begin{figure}
\begin{center}
\rotatebox{0}{
\epsscale{0.90}
\plotone{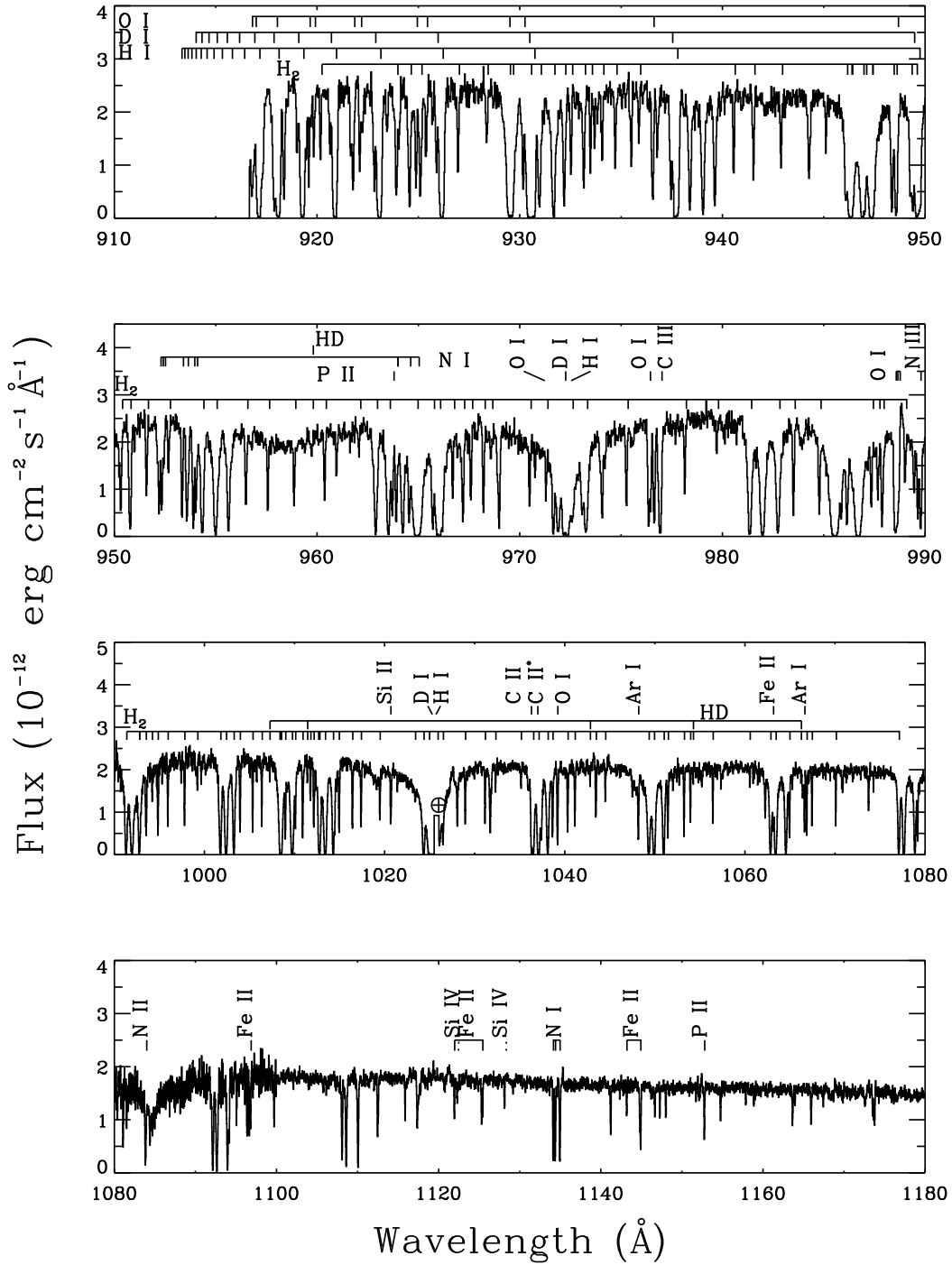}}
\caption{\FUSE\ data  for \pgoo\ with ISM and geo-coronal ($\oplus$)
identifications.  Data are binned by four pixels, scaled and merged for
presentation only. Photospheric lines of \ion{Si}{4} 
are marked with dashed lines. We could not
identify some absorption lines, particularly in the range 
1080--1180~\AA\ (Table~\ref{tab-unidentifiedlines}; a
stellar origin is suspected). 
\label{fig-totalspectrum}}
\end{center}
\end{figure}

The \FUSE\ data were processed with the CalFUSE  2.2.2 pipeline. We summed the \FUSE\ data
interactively using software provided by S. Friedman, {\sc corrcal}, registering each exposure based on
cross-correlations around narrow absorption features near the middle of each channel segment.
See 
Fig.~\ref{fig-totalspectrum} for the full spectrum.  We then measured and corrected the $\sigma$ error array and
wavelength calibration to prepare the data for profile fitting.  See Appendix~\ref{appendix-reduction} for details.

\section{Analysis}
\label{sec-analysis}

We prepared a grid of stellar models to provide a first estimate for the continuum for
the \FUSE\ data. We then used a suite of techniques, software and analyists  to analyze
the data, in an attempt to limit errors dependent on individual algorithms, programs or
persons.  In particular, we  employ curve-of-growth (COG), apparent optical depth (AOD)
and profile fitting (PF) techniques at various points in the analysis, and present
results both from the profile fitting codes \vpfit\ \citep[][see Bob Carswell's site
http://www.ast.cam.ac.uk/$\sim$rfc/]{Webb87} and \owens\ \citep{Lemoine02} obtained by
three different co-authors: C.~Oliveira (CO) and 
G.~H\'ebrard (GH) for \owens\ and G.~Williger (GW) for \vpfit.  We stress that
all analysis methods used the same data and error values.

\subsection{Stellar model grid}
\label{sec-stellarmodelpg0038}

To determine the effect of uncertainties in the continuum from 
the stellar model on column densities of ISM species,
a grid of models was computed using TLUSTY \citep{Hubeny95}
around the best-fit values for effective temperature and surface gravity
$T_{eff} = 115000~\rm{K}\pm 11500$ and $\log g = 7.5\pm 0.3$
\citep{Dreizler96}.  We fixed
[He/H] to 1000, because no photospheric \HI\ absorption is seen
in the \FUSE\ data; the precise H abundance is difficult to constrain due to the strong ISM absorption
and blending with \ion{He}{2} absorption.  The effect of any uncertainty on higher order \DI\ lines 
for such a low \HI\ abundance should be negligible.
The models included nine energy levels for \HI\ and twenty for \ion{He}{2}, plus line blanketing 
caused by them.

Synthetic spectra containing only H and He for simplicity
were computed over the \FUSE\ range using the program SYNSPEC
\citep{Hubeny95} and convolved with the instrumental profile using the program ROTIN3
(provided with SYNSPEC). We find photospheric lines in the data from 
\ion{N}{4}~$\lambda$~923.7, 924.3, 
\ion{O}{6}~$\lambda$~1031.9, 1037.6, 
\ion{Si}{4}~$\lambda$~1128.3, 
\ion{Si}{5}~$\lambda$~1117.8 (blended with \ion{Fe}{7}),
\ion{S}{6}~$\lambda$~1117.8 and
\ion{Fe}{7}~$\lambda$~1117.6, 1163.9, 1164.2, 1166.2 (and possibly 1146.5, 1155.0).
We also computed a model with metals in addition to H and He to investigate
the influence of metal lines on the Ly$\beta$ profile (\S4.6), 
with $T=115000$~K and $\log g =
7.5$, [He/H]=5. (a conservative lower bound), 
[C/H]$=5\times 10^{-5}$, [N/H]$=5\times 10^{-3}$ and [O/H]$=5\times
10^{-5}$, which is compatible  with the \citet{Dreizler97} upper limit on
hydrogen abundance (not formally detected).   

\begin{figure}
\epsscale{1.05}
\plotone{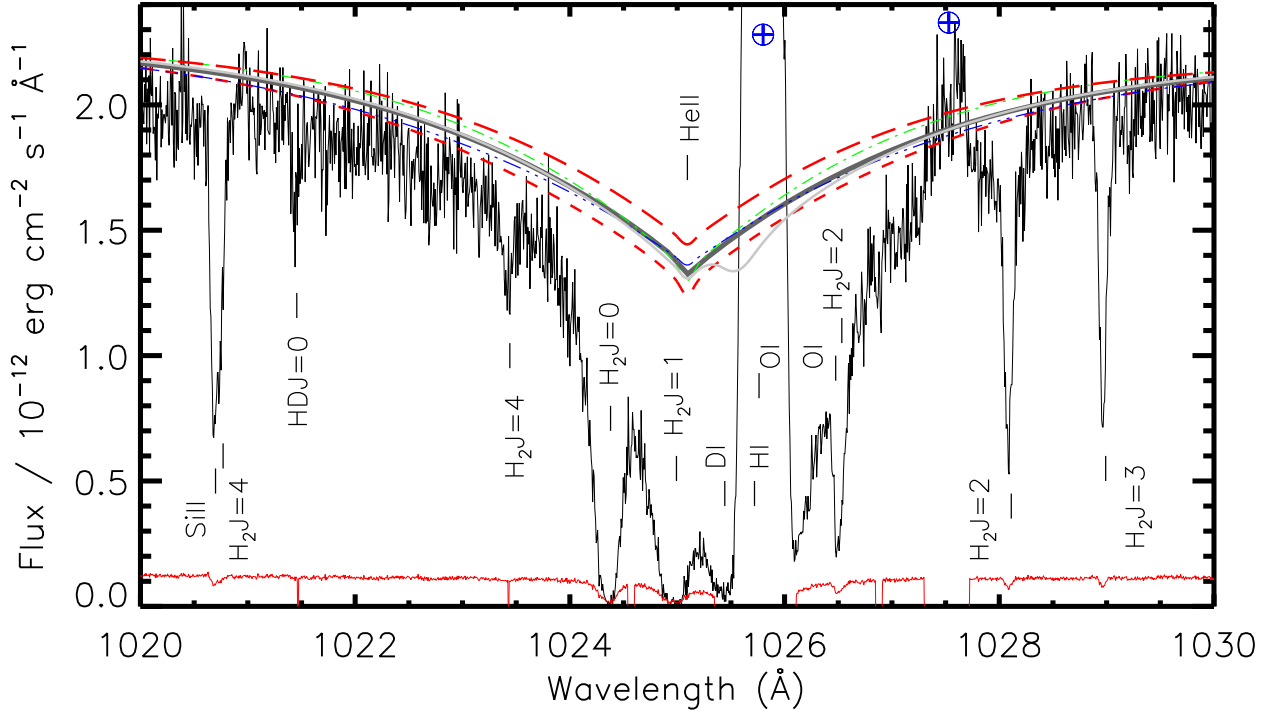}
\caption{The grid of stellar models described in \S~\ref{sec-stellarmodelpg0038}, 
shifted by radial velocity $v=-52$~\kms\ and
plotted against the LiF1A data and $1\sigma$ errors  around \HI\ \Lyb .
{\it Thick (dark grey) solid line:} $T=115000$~K, $\log g = 7.5$, He-dominated (reference model).
{\it Thick (red) dashed lines:} $T=103500$~K, $\log g = 7.5$, He-dominated (lower) and 
$T=126500$~K, $\log g = 7.5$ (upper).
{\it Thin (green) dash-dotted line:} $T=115000$~K, $\log g = 7.2$, He-dominated.
{\it Thin (blue) dash-triple dotted line:} $T=115000$~K, $\log g = 7.8$, He-dominated.
{\it Thin (light grey) solid line:} $T=115000$~K, $\log g = 7.5$, 
[H/He]=0.2, [C/H]$=5\times 10^{-5}$,
[N/H]$=10^{-3}$ and [O/H]$=5\times 10^{-5}$.  $\bigoplus$: Geo-coronal \Lyb , \oi $\lambda\, 1027.5$ emission.
ISM and photospheric absorption lines are noted with ticks and labelled.
The feature at 1022.7~\AA\ is not in the LiF2B data and appears to be noise.
\label{fig-stellarmodels}
}
\end{figure}

For the purposes of fitting a stellar continuum to the data, the
stellar
radial velocity is estimated to be $-52$~\kms , based on \ion{O}{6} photospheric absorption lines.
We then fitted the model
spectra described above to the \FUSE\ data,
using separate  flux normalizations for each of the eight \FUSE\ 
detector/channel
segments.  
The models are shown for the \Lyb\ region in Fig.~\ref{fig-stellarmodels}.

The stellar models were used as a first attempt at a continuum fit for absorption
line analysis, except for LiF1B, which had to have a continuum fitted by hand due to the
``worm" feature \citep{Sahnow00}, an optical anomaly due to the astigmatism
in the system and a wire mesh in front of the detector.  
The stellar models were finally given a minor 
linear correction, determined by eye, to match the flux level in each channel segment.
Further local zero-point corrections to the continuum will be allowed in the subsequent analysis.
We see no indication of scattered light.

\subsection{Methods of line analysis}
\label{sec-lineanalysismethods}

{\it Curve of growth:} The COG technique is often used with \FUSE\ data because 
it offers an independent technique to determine column densities and Doppler parameters
for species with many different oscillator strengths, such as \HH .  See \citet{Oliveira03}
for details.

{\it Apparent optical depth:} In the AOD technique, the column density is determined by directly
integrating the apparent column density profile over the velocity
range of the absorption profile.  See \citet{Savage91} and \citet{Sembach92} for details.
We use the AOD technique as a consistency check of the results obtained
with PF and COG, and to determine lower limits on the column densities of
species which only have saturated transitions in the $FUSE$~bandpass.

{\it Profile fitting:}  If there are several components contributing to a given
atomic transition, profile fitting in principle yields much more  information than
simple curve of growth analyses. However, the \fuse\ resolution of $15-20$~\kms\ is
not high enough to resolve typical ISM lines, which generally arise from a number of
narrow components over a small velocity interval.  \HI\ lines are particularly
challenging, in that their intrinsic widths are high due to their low atomic mass. 
Nevertheless, for blended complexes of lines, one can obtain reasonably accurate
estimates for total column density (with a corresponding inflated composite Doppler
parameter), provided that the distribution of intrinsic component optical depths and
Doppler parameters is smooth \citep{Jenkins86}.  We employ two profile fitting
programs for this work: \Owens\ and \vpfit .  \Owens\ was developed by M. Lemoine and
the French  \fuse\ team at the Institut d'Astrophysique de Paris, and has been used
for a number of \DHrat\ studies of Galactic targets involving \FUSE\ data.  See
\citet[][]{Lemoine02,Hebrard02,Hebrard03,Oliveira03} for a fuller description.  
\vpfit\ has been used for many years, but its contribution to \DHrat\ work has  so
far concentrated on extragalactic sight lines
\citep[e.g.][]{Webb91,Carswell96,Pettini01,Crighton03}.   We use both codes here as a
check for the robustness of results between the two.

\vpfit\  
originally was developed for the analysis of ground-based echelle spectroscopy, but has also
been used on various high resolution \HST\ spectra.  It generally works best with well-defined
wavelength solutions, line spread functions and error arrays, which can be a challenge for
\FUSE\ data, and normally is not used to work with as many
fitting windows as \owens .  
\vpfit\ also requires more steps than \Owens\ to deconvolve the effects of turbulent and thermal
velocity.  Nevertheless, \vpfit\ is easily adaptable and well-supported.
Its strength is that it can easily accommodate highly blended complexes, and can
generate a $\chi^2$ grid relatively quickly compared to \owens . 
For \vpfit\ results, 
our errors are calculated based on $\Delta \chi^2=\chi^2_{min}+1$ for
$1\sigma$
and $\Delta \chi^2=\chi^2_{min}+4$ for $2\sigma$. 
Like \owens , \vpfit\ can allow a wavelength offset for each fitting region to accommodate
uncertainties in the dispersion solution.  
Similarly, \vpfit\ can permit variations in continuum level.  We leave such variations
as zero point offsets in those windows which require them, because it is unlikely (and not
desirable) that the continuum slope depart significantly from that of the stellar model
over the short range (on the order of $\simlt 1$~\AA ) of each fitting window.
The reliance of \vpfit\ on a pre-determined continuum with zeroth order adjustments is 
one of its main differences from \Owens , which uses polynomial approximations 
to model flips and dips in the local continuum and, if desired, to approximate absorption line wings.
We employ \vpfit\ as a check
for general \owens\ results and the binning used in its analysis, and typically use \vpfit\ to 
solve for column densities and Doppler parameters
for one or a small number of species at a time.  For this reason, we generally leave the Doppler
parameter free to vary by species, with the benefit that unphysical Doppler parameters may signal
that the LSF or blended component structure may be poorly understood in a fitting region.

For the \vpfit\ analysis, the data were left unbinned, to permit the exclusion of
the narrowest intervals which
are suspected noise spikes and regions 
contaminated by airglow.  For the AOD, COG and \Owens\ analyses in this paper, the data were chosen to be 
binned by three pixels to increase the signal to noise
ratio, as allowed by the oversampling of the LSF.
For the \vpfit\ analysis, we
use negative $\sigma $ values as flags to omit data from profile fits, whereas for \Owens ,
bad data were flagged by high $\sigma$ values.
Wavelength ranges for measurement intervals 
for the COG, AOD and the two PF methods were not required to be the same.

All three methods described above provide formal statistical error estimates.
Systematic errors, which were discussed in detail by \citet{Hebrard02},
also have an important effect in overall error estimates, and will also be considered 
in this study.

\subsection{\NaI\ Optical Data}
\begin{figure}
\begin{center}
\rotatebox{0}{
\epsscale{0.85}
\plotone{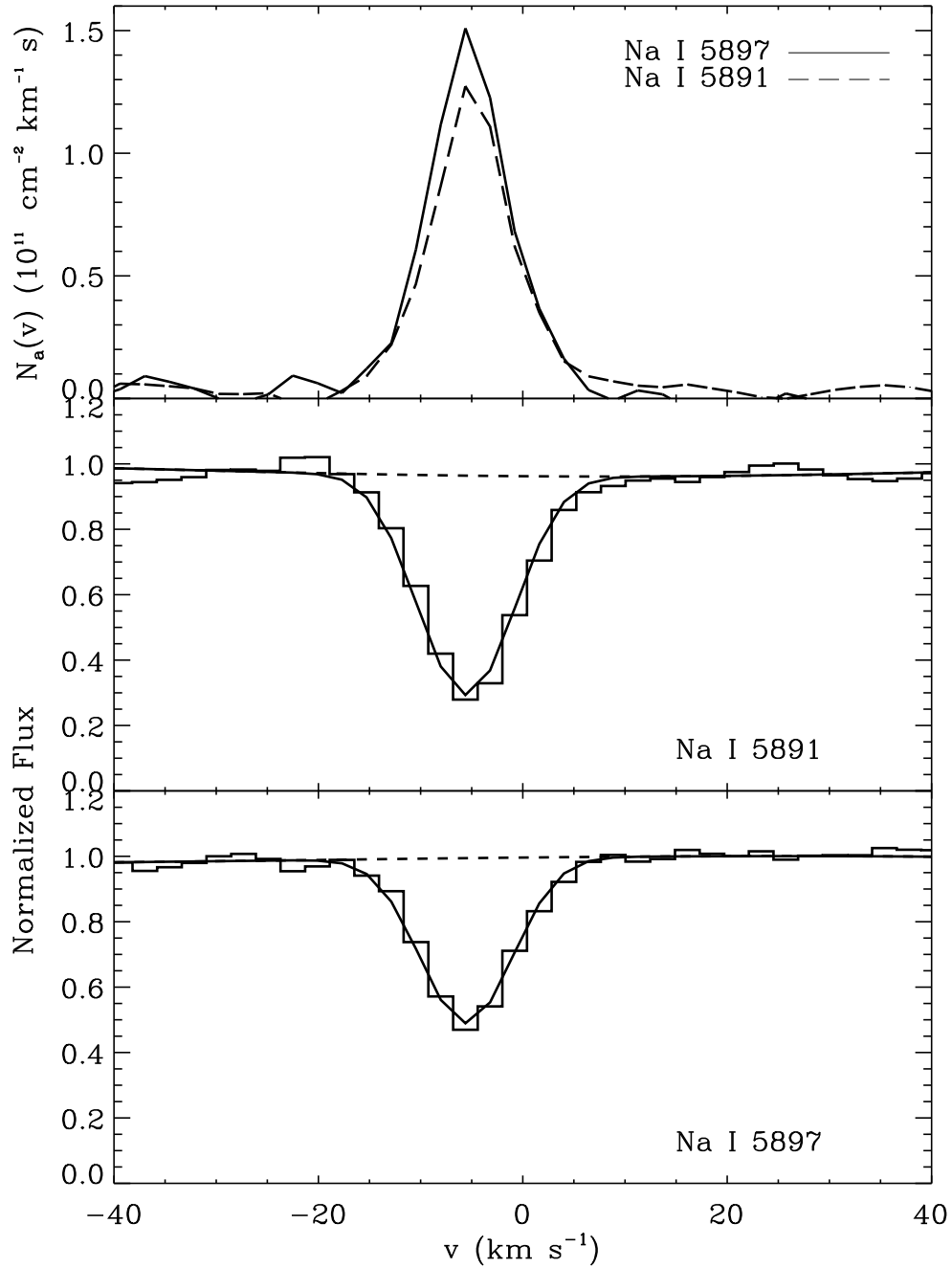}}
\caption{{\it Top panel:} Apparent column density per velocity unit for the \NaI\ doublet
observed toward \pgoo . The $f$-values of the two transitions differ by
a factor of 2. A small amount of unresolved saturation can be seen in the range
$-12 < v < -1$~\kms\, where the profile of the stronger line
falls below that of the weaker line. {\it Bottom two panels:} \Owens\ fit to the Keck HIRES
data.
\label{fig-NaI_AOD}}
\end{center}
\end{figure}

The Keck spectra of the \NaI~5891,5897 lines are shown in Fig.~\ref{fig-NaI_AOD}.
We used the PF and AOD techniques to determine the column density of \NaI . 
Following the procedure
outlined in \citet{Sembach92}, 
we find equivalent widths W$_{\lambda}$($\lambda$5891) = 173.72
$\pm$ 5.62 m\AA, and W$_{\lambda}$($\lambda$5897) = 107.71 $\pm$ 4.00 m\AA .

\subsubsection{Profile Fitting}

Profile fitting of the \NaI\ doublet with \owens\ was performed independently of the other
atomic species, with one and two absorption components. The column densities
derived with one and two components are consistent with each other: $\log N(\NaI)=12.19\pm0.04$ ($2\sigma $). 
The fits are
shown in the bottom panels of Fig.~\ref{fig-NaI_AOD}.
Using two
components does not significantly improve the quality of the fit determined
from the \chisq\ statistic, leading us to conclude that 
at the Keck spectrum's resolution of 8.3 \kms\ the 
\NaI\ lines are consistent with a single absorption component centered at 
$v\sim -$6~\kms . (Radial velocities are given in the heliocentric frame).  
\vpfit\ gave a satisfactory
fit for a single component at
$v = -5.1\pm 0.3$~\kms , 
with Doppler parameter $3.6^{+0.4}_{-0.2}$~\kms\  and
$\log N(\NaI)=12.25\pm 0.04$ ($2\sigma $ errors).
We adopt $v=-5.1$~\kms\ as an absolute velocity reference for the neutral
gas in the \FUSE\ data.
We note that \NaI\ is an imperfect tracer of \HI\ \citep{Cha00}, but
have no other high
resolution information for the ISM absorption velocity structure.

\subsubsection{Apparent Optical Depth}

The top panel of Fig.~\ref{fig-NaI_AOD} compares the apparent column density of the
two components of the \NaI\ doublet as a function of
velocity. The
two profiles are not consistent in the range $-12 < v <-$1~\kms ,
indicating that there is some unresolved saturated structure at these velocities. 
Disagreements between the two profiles for $v <-20$~\kms\ and $v > +4$~\kms\ are at
the level of noise in these regions. We derive apparent column
densities, $\log N(\NaI ) = 12.15 \pm 0.02$ for the weaker line (5897.56 \AA) and
$\log N(\NaI ) = 12.11 \pm 0.02$  for the stronger line (5891.58 \AA).
Since the
difference between the log of the two apparent column densities is less than 0.05~dex,
we can use the procedure outlined by \citet{Savage91} to correct for the small
saturation effects in these transitions.  A better
estimate of the true column density is then achieved with  $\log N_{a} = \log
N(\NaI )_{\lambda 5897}+\Delta\log N(\NaI )_{\lambda 5891-5897}$,  where the first term in the equation is the
column density derived from the weaker member of the doublet (\NaI~5897~\AA ) and the second term
corresponds to the difference in column density derived from the two lines of the
doublet. Using the equation above we obtain 
$\log N(\NaI ) = 12.19 \pm 0.04$ ($1\sigma$), 
consistent with the profile fitting results and which we adopt.

\subsection{Molecular hydrogen}
\label{sec-h2pg0038}

The \pgoo\ \FUSE\ data have a rich \HH\ absorption spectrum, and
many of the \HH\ features are saturated and/or blended.
It is crucial, however, to have a good constraint on the various \HH\ transitions,
because the lines are often blended with 
\DI , \HI\ and all other atomic ISM species.  We used both curve-of-growth and
profile fitting methods to constrain $N(\HH )$.

\subsubsection{Curve of Growth}

\begin{figure}
\begin{center}
\rotatebox{90}{
\epsscale{0.8}
\plotone{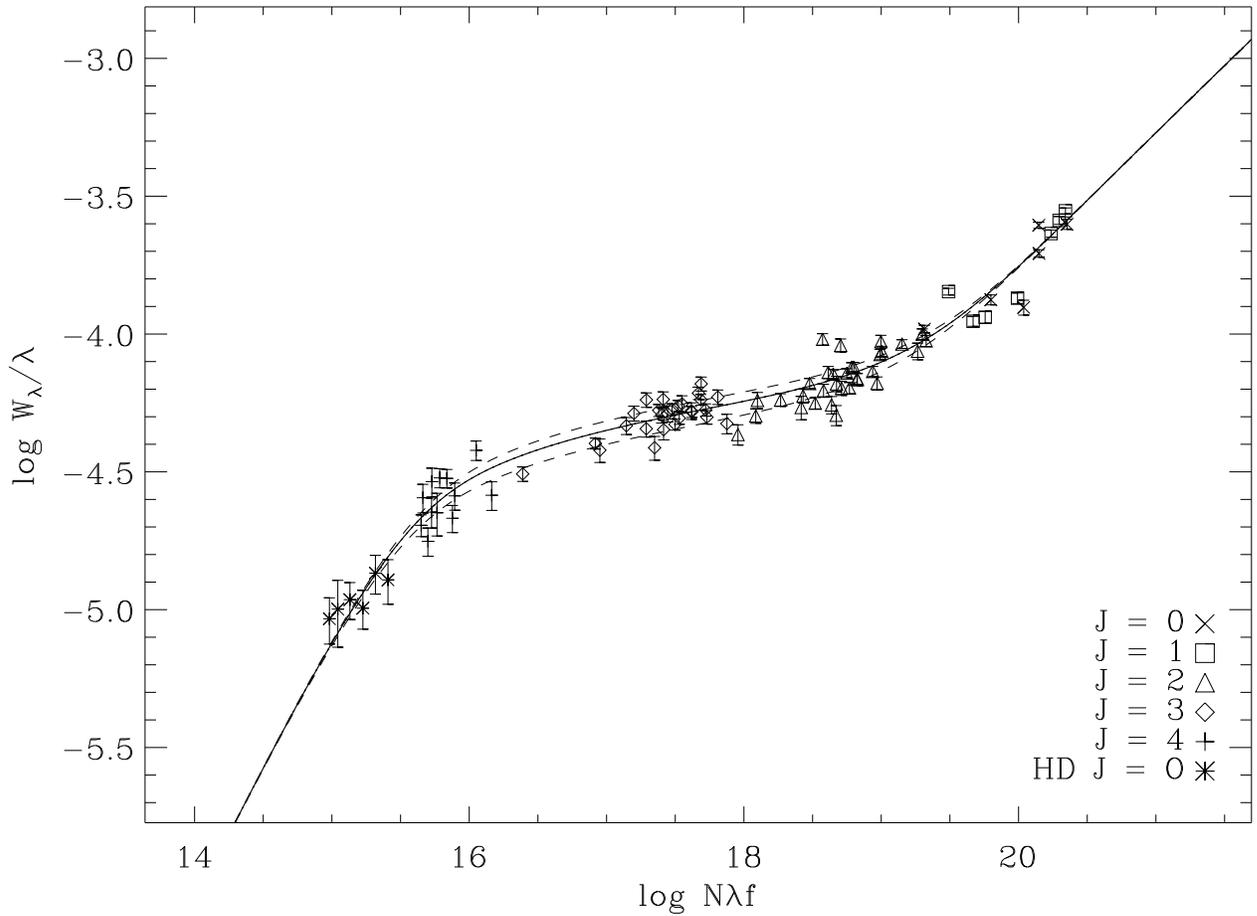}
}
\caption[COG for \HH\ and HD along the \pgoo\
sight line]{Single-component COG for \HH\ and HD along the \pgoo\
sight line. A different symbol is used for each $J$ level. 
The $J=5$ level was not included because the equivalent widths for its
transitions are too small.
\label{fig-pg0038h2cog}}
\end{center}
\end{figure}

For the COG method it was assumed that a single Doppler parameter, $b$,
describes all the rotational levels of \HH\ and HD. For both of these
molecules, equivalent widths corresponding to a total of 
91 different transitions (6, 7,
33, 25, and 14 for the $J = 0, 1, 2, 3, 4$ rotational levels of
\HH\ respectively; 6 for HD $J$ = 0) were measured in all the \fuse\ channels
where they are detected (Table~\ref{tab-h2hdlines}). We did not use $J=5$ for a COG analysis because it is too weak.
The equivalent widths for each line, measured in different
detector/channel segments, were then compared. All of them agree within the 2$\sigma$ errors, though we
regard the results for $J=0$ and 1 with caution due to blending. These
equivalent widths were then combined through a weighted average, and a
single-component COG  was fitted to the 91
values. It is shown in Figure~\ref{fig-pg0038h2cog}.
The COG fit yields $b = 3.1^{+0.3}_{-0.2}$~\kms .

\subsubsection{Profile Fitting}
\label{subsec-H2profilefit}

For profile fitting with \owens , CO used a single Gaussian line spread function (LSF) 
to describe the instrumental line spread
function, with a full width FWHM of 10.5 unbinned pixels.
GH allowed the LSF to vary. 
\citet{Hebrard02} found from \owens\ profile fits for \DHrat\ work
that $\Delta \chi^2=4$ (or $2^2$)
at the $2\sigma$ level for errors in line fitting parameters such as
column density, $\Delta \chi^2=9$ (or $3^2$) for $3\sigma$, and
similarly up to $7\sigma$. 
This $\Delta \chi^2=k^2$ for $k\sigma$ relation is used in
our \owens\ analysis to establish $2\sigma$ errors based on averages from 1$\sigma$ to
7$\sigma$.
The profile fitting of the \HH\ and HD
lines was performed simultaneously with the other atomic species for which the
column densities were being sought. We used 11, 18, 12, 8, 18, 14 lines for
the $J=0, 1, 2, 3, 4, 5$ \HH\ rotational levels in all the \fuse\ channel segments where they
were seen.
\HH\ and HD were not restricted in radial velocity in comparison to atomic species
nor, in the case of the analysis of GH, to each other, who found 
a radial velocity difference $|v_{\mathrm H_2}-v_{\mathrm HD}|=4.6$~\kms\ using this approach.
We derive $b_{\mathrm H_2 , HD}\sim 2.7$~\kms .

\vpfit\ is less well-suited than \owens\ for fitting simultaneously the very
large number of spectral regions containing \HH\ and HD lines  in the \pgoo\
data.   We therefore used it to determine the wavelength-dependent LSF, based on
\HH\ column densities and Doppler parameters from the COG and \Owens\ results
(Appendix~\ref{appendix-h2lsf}). It is then possible to offer a consistency
check and estimate of the profile fitting  uncertainties with \vpfit , using a
subsample of fitting regions from the LiF1A segment.  We made trial fits fixing
the Doppler parameter at $b=2.7$,~3.1,~3.4~\kms\ based on the PF and COG
estimates, and took the range of \HH\ column densities as a function of $b$ for
the various rotational levels as roughly $2\sigma$ systematic errors.   The
range in values from the COG and \Owens\ PF measurements were also treated as 
$2\sigma$ contributions to the systematic error. We then 
compared the $2\sigma$ 
statistical errors for the various PF and COG methods, and the variations
in column density as a function of Doppler parameter and the variations
in column density between the various PF and COG methods as approximate
$2\sigma$ errors in the approach taken by \citet{Wood04}, to form
overall $2\sigma$ estimates in the \HH\ and HD column densities.
The
exact \HH\ column densities do not make a significant difference to atomic column
densities except in the case of \HI\ (\S~\ref{sec-hipg0038}).

\HH\ and HD column densities are presented in Table~\ref{tab-pg0038h2columns}. 
The quoted uncertainties are $2\sigma$.  Discrepancies in the COG and
PF results make $1\sigma$ uncertainties difficult to estimate, but we
assume that $1\sigma$ errors would be roughly half the $2\sigma$ errors.
The total \HH\ column density 
along this line of sight is $\log {\mathrm N}(\HH )  = 19.33\pm
0.04$ ($2\sigma$),  and $\log ({\mathrm {HD}}/\HH ) = -5.38^{+0.06}_{-0.05}$ ($2\sigma$).  Column
density results were insensitive to the exact LSF used.

\begin{figure}
\begin{center}
\rotatebox{90}{
\epsscale{0.8}
\plotone{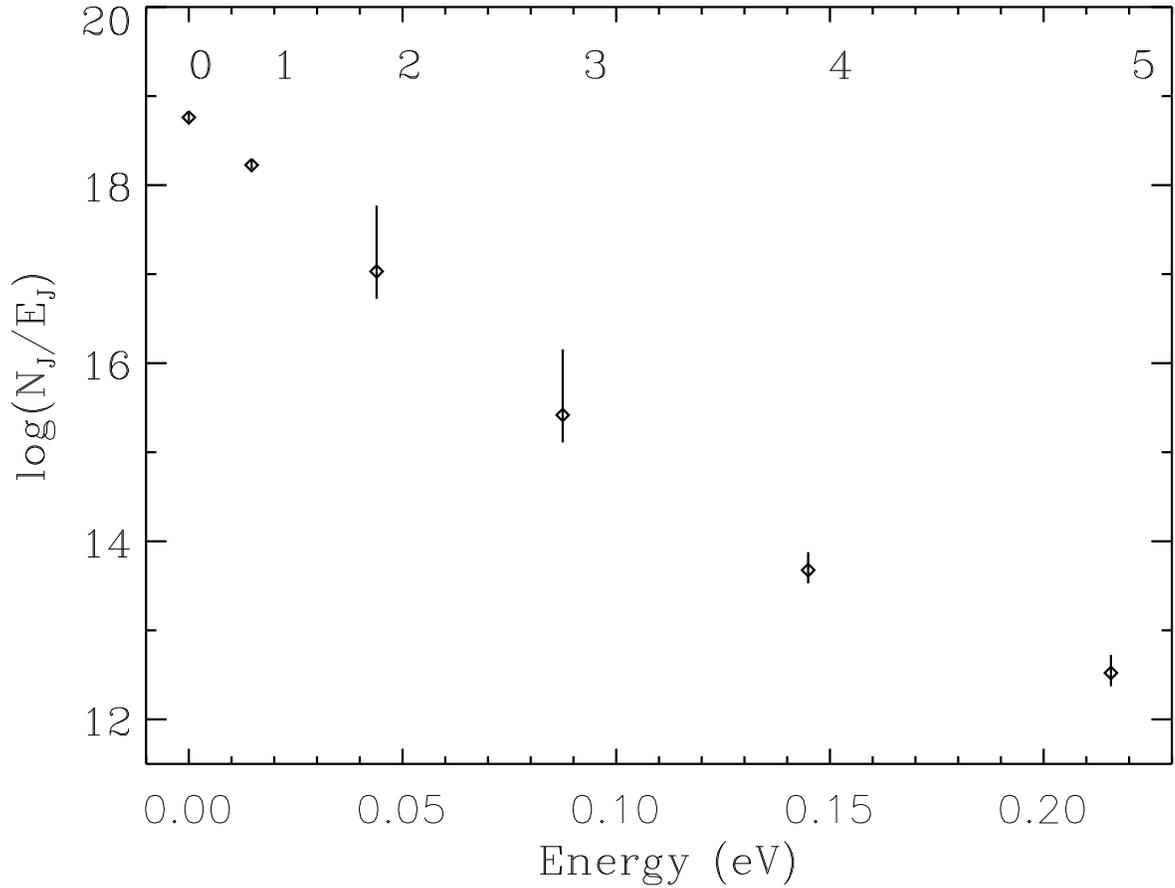}
}
\caption[Excitation for \HH\ along the \pgoo\  line of
sight]{Excitation for \HH\ along the \pgoo\  line of sight,
using the adopted values from Table~\ref{tab-pg0038h2columns}. \HH\ J levels are
marked at top.  See text
for details.
\label{fig-pg0038h2exc}}
\end{center}
\end{figure}

To estimate the \HH\ excitation temperature, we fitted the $J = 0$ and 1 levels
independently of the other levels, and find $T_{\rm 01} = 136 \pm 21$~K.
The column densities for the various $J$ levels 
divided by their
relative statistical weights are shown in 
Fig.~\ref{fig-pg0038h2exc}.   
We can also fit the \HH\ $0<J<4$ and HD $J=0$ column densities with a single
temperature, $T_{\rm 04} = 143 \pm 5$~K.  
We note that 
the $0<J<4$ levels are not usually consistent with a single temperature,
and that $T\sim 140$~K is unusually high for Galactic \HH\
\citep{Black73,Snow00,Rachford01,Rachford02}.
Using the \HI\ column
density derived in \S~\ref{sec-hipg0038}, we compute the fraction of \HH\ along
the sight line,  $f(\HH ) = 2 N(\HH )/(N(\HI ) + 2N(\HH )) = 0.14\pm 0.02$
($2\sigma$ errors).

Finally, we checked the accuracy of our continuum placement by examining 
33 HD lines in the six FUSE segments covering
915-1100~\AA .  The HD lines are optically thin, and therefore quite
sensitive to continuum level.
\vpfit\ can calculate zero point continuum correction
factors for each of the 33 fitting windows.  
A simultaneous fit to the HD lines
yields an average correction factor of
$\langle {\rm cont} \rangle = 0.967\pm 0.044$.  The
continuum we use is thus not systematically low or high, and adjustments
have a scatter of 
$\sim$4\%.

\subsection{Column Densities of Metal Species}
\label{sec-atomicpg0038}

The column densities presented in this work for the atomic species were
determined with the PF, AOD and COG techniques.  
Wavelengths and $f\lambda$ values
(and equivalent widths for \FeII )
of the metal, \HI\ and \DI\ transitions used in this work
are listed in Table~\ref{tab-atomiclines}.
In some cases, it was only possible to quote a lower limit via the AOD method 
for column densities
due to 
saturated transitions in the \fuse\ bandpass. 
The COG was used only with \FeII , because
this species is the only one which has enough non-saturated and non-blended
transitions sufficient for this method.

We included in our model fit for metal lines a molecular
cloud with \HH\ and HD, as described in \S~\ref{sec-h2pg0038}, because  many
atomic transitions are blended with \HH .  
The \CI\ absorption will
follow cool gas like that responsible for the \HH\ and \NaI\ absorption
rather than the warm gas that will be responsible for much of the
\HI , \DI , \NI\ and \OI\ absorption.  Therefore,
for \owens\ PF results, for co-author CO,
\DI , \NI , \OI\ and \FeII\ were fitted in one component, and
\CI\ and \HH\ in another.  Co-author GH did not include \CI\ in his fits.
Atomic column densities are listed in
Table~\ref{tab-pg0038atomiccolumns}.

An example fit to \OI~$\lambda 974$,
which is the only available unsaturated \OI\ line and the entire basis of our \OI\ column
density measurement, is in Fig.~\ref{fig-traceOI}.  It is strongly blended with
\HH\ $J=2$ and 5.  The resolution in SiC2A at 974~\AA\ is $\sim 18$~\kms , which is among the
best in our data.  However, in SiC1B it is $\sim 30$~\kms , among the worst.
The SiC2A data therefore provide the strongest constraints.  Co-author GH only
used the SiC2A data for his \Owens\ fit, while co-authors CO and GW also included the
SiC1B data for their \Owens\ and \vpfit\ (see below) calculations, respectively.  All three co-authors
found consistent results with each other.
We regard $N(\OI )$ with caution, because of its
dependence on one line with good resolution only in the SiC2A data.
We will discuss $N(\OI )$ further in \S~\ref{subsec-columndensityratios}.

For \vpfit\ PF results, we used the wavelength-dependent LSF derived in
Appendix~\ref{appendix-h2lsf}, and fitted each species independently.   The
Doppler parameters were left free to vary, as a check for LSF problems or
evidence of substantial unresolved blending.   A significant ($>2\sigma$)
difference between \vpfit\ and \Owens\ results only arose for \NI , for which
the best fit Doppler parameter would require 
$b=9.0^{+1.1}_{-0.5}$~\kms , which is high compared to $b$ from \HH .   
If we fix the Doppler parameter to $b_{\mathrm
{NI}}=3.1$~\kms\ (based on the value from \HH ), then we obtain $N(\NI )$ from
\vpfit , consistent with  the fits from both co-authors who used  \owens . 
However, despite the simplicity of the \NaI\ absorption,
it is likely that the \HI , \OI\ and \NI\ absorption (which will be from
warm as well as cold gas) will have multiple components, in which case
$b=9.0$~\kms\ is {\it not} unreasonably high but is simply a consequence of the
unresolved velocity structure.  Likewise, assuming the $b$ value from
the \HH\ fit for \NI\ may be completely inappropriate since to a large
extent the \HH\ and \NI\ absorption may be coming from entirely different
places.

Otherwise, the results are consistent between the one \vpfit\ and two \Owens\
calculations, and in particular are independent of how \CI\ is included in the
fits. The column densities derived for the atomic species are presented in 
Table~\ref{tab-pg0038atomiccolumns}, and are a compromise of \owens\ and
\vpfit\ values. We also found a number of unidentified lines,  listed in
Table~\ref{tab-unidentifiedlines}, which may be either stellar or interstellar.

\subsection{\HI\ and \DI\ Column Densities}
\label{sec-hipg0038}

The \Lya\ transition provides the best determination
of $N(\HI )$ for the highest column densities, 
based on damping wings. 
Unfortunately, only \IUE\ low resolution data
cover the \Lya\ region for \pgoo , and the ISM \Lya\ trough is filled in, presumably by
geo-coronal emission.  
\FUSE\ provides wavelength coverage from the Lyman limit up to \Lyb .
We must therefore make an estimate of \hi\ column density
based on \Lyb , which is
blended with a large number of \HH\ and \oi\ 
transitions, but which offers promise due to strong damping wings.

\subsubsection{\HI\ and \DI\ column densities}

{\it H~I:} For the \Owens\ calculations, we fitted the \lyb\ profile separately 
from the other species.
However, because \HI\ and \DI\ are blended with  \oi\ and \HH , we used the column
densities of the latter, determined in \S~\ref{subsec-H2profilefit}--\ref{sec-atomicpg0038}, as fixed
constraints. All four of the \HI\ \lyb\ profiles obtained with the
LiF and SiC channels were used simultaneously.  The two LiF channels dominate the results because of their
higher signal to noise ratio.

\begin{figure}
\begin{center}
\rotatebox{0}{
\epsscale{0.85}
\plotone{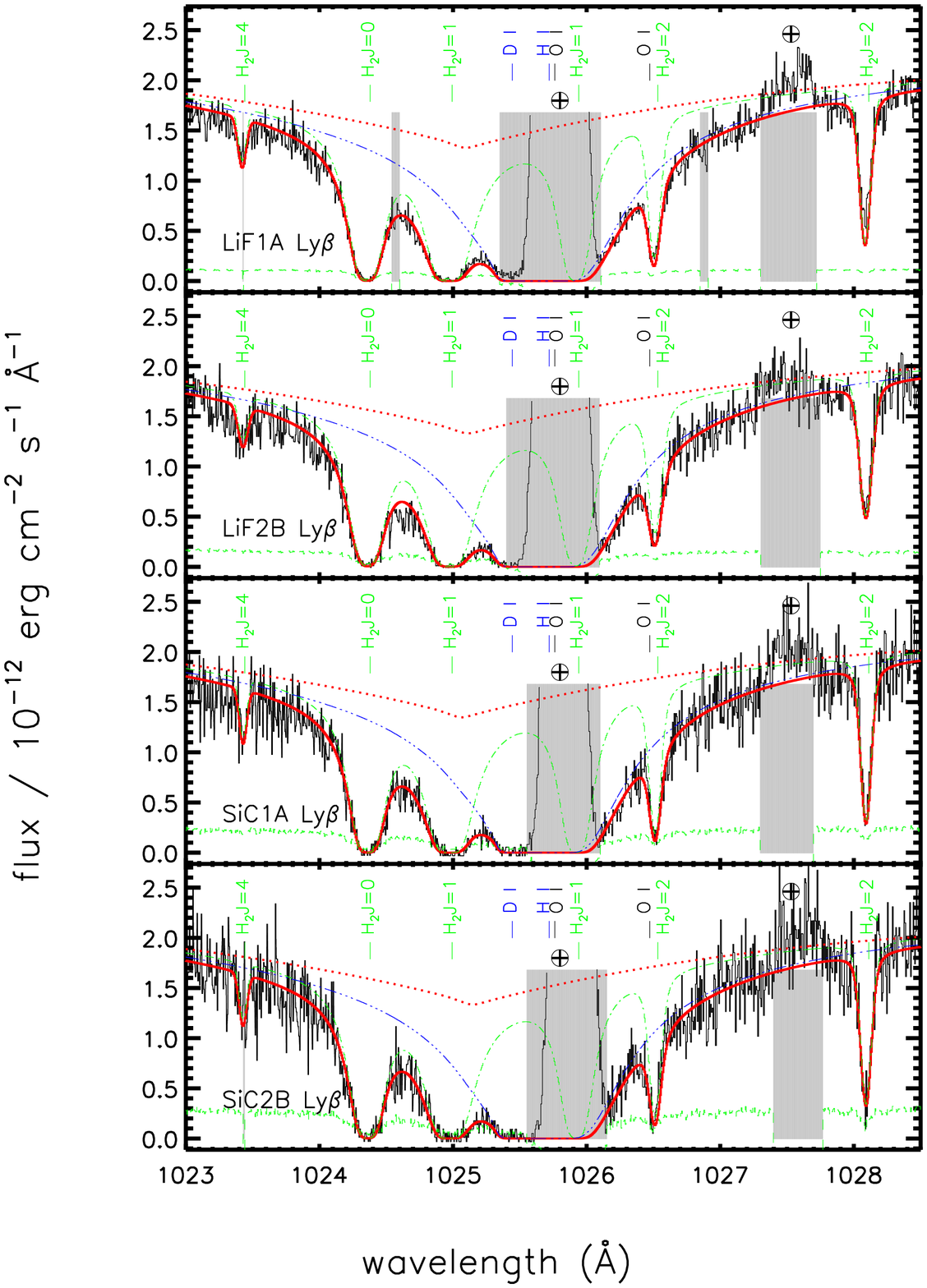}
}
\caption[\vpfit\ fit to \HI\lyb~profile]{\small
{\it Solid (red) line}: \vpfit\ combined fit to \hi , \di , \OI\ and \HH . 
{\it Dashed (green) line}: $1\sigma$ error array.
{\it Dash-dotted (green) line}: \HH\ profiles (all lines shown after convolution with
local LSF).
{\it Dash-triple-dotted (blue) line}: HI, DI profiles.
{\it Dotted (red) line:} Stellar continuum.
\OI\ profiles are not shown because they fall in geo-coronal \Lyb\ emission and are
dominated by \HH\ lines.
$\bigoplus$: Geo-coronal \Lyb , \oi $\lambda\, 1027.5$ emission.
{\it Grey zones}: regions which were 
excluded from the fit, including geo-coronal emission and noise spikes.  
\HH , \hi , \di\ and \oi\ transitions are
marked by ticks.  
\label{fig-HIvpfitpg0038}}
\end{center}
\end{figure}

The \vpfit\ analysis of the \hi\ column density included a simultaneous fit
to all segments covering \lyb , Ly7, Ly8, Ly9 and Ly10.  However only \Lyb\ determines
the column density, with the profile fit shown in Fig.~\ref{fig-HIvpfitpg0038}.  
The higher order lines were included in the fit for consistency and
to constrain the Doppler parameter, as will be discussed below.
We did not include Ly$\gamma$, $\delta$, $\epsilon$ and Ly-6 in the fits because
they do not present strong enough damping wings to constrain the \HI\ column density,
and are too saturated to constrain the \DI\ column density.
As a check, we did obtain a statistically consistent fit for our adopted results (see below)
for Ly$\gamma$ and Ly$\delta$ with \vpfit , after exercising care in excluding regions
affected by geo-coronal emission, which we confirmed by separating our data into
day and night subsets.\footnote{Geocoronal emission for Ly$\gamma$ has a weak red wing.
Upon consultation with D. Sahnow and P. Feldman (2004, private communication), a similar effect
is seen in Ly$\delta$ and Ly$\epsilon$ emission.  This is most likely
an instrumental effect arising from 
geocoronal illumination of the entire LWRS aperture.}
The various \HH\ and HD column densities 
were fixed according to the values in Table~\ref{tab-pg0038h2columns}, which
resulted in a best fit of
$\log N(\HI ) $ = 20.41.  Errors will be discussed in the next section.
The effective \HI\ Doppler
parameter is $13.6\pm 0.2$~\kms\ ($2\sigma$ errors), 
which is most likely an inflated value of $b$ due to the
presence of unresolved components \citep{Jenkins86}.

\begin{figure}
\begin{center}
\rotatebox{0}{
\epsscale{1.05}
\plotone{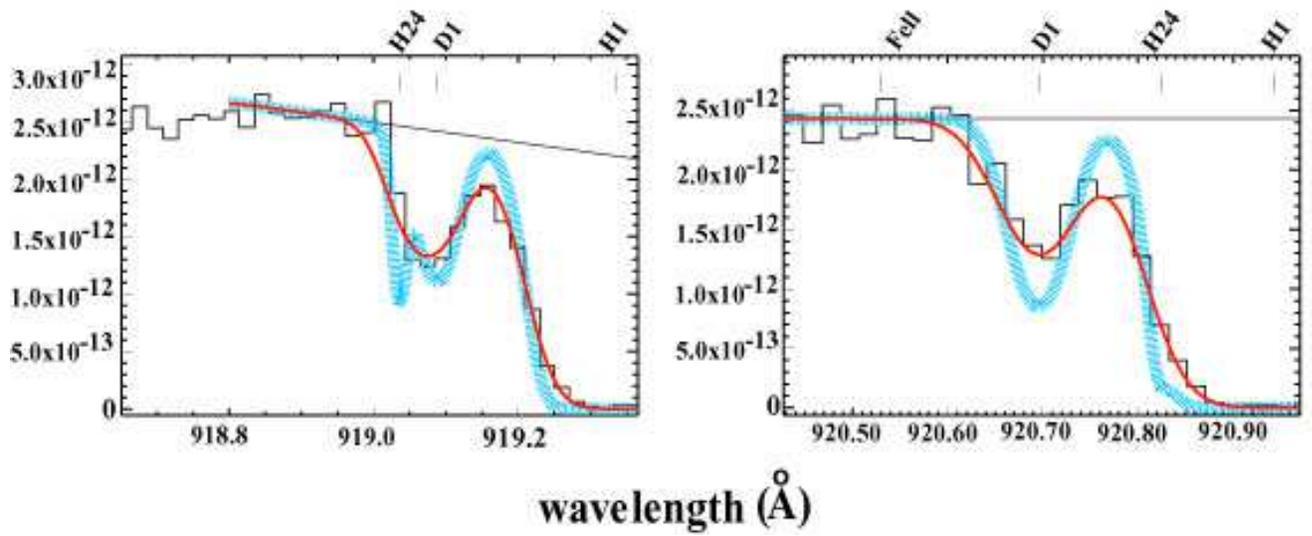}
}
\caption[\Owens\ fit to \DI\ Ly9 and Ly10 profiles]{\Owens\ fit to \DI\ Ly9 and Ly10 profiles
(SiC2A channel, normalized data, binned by 3 pixels). {\it Red lines:}  fit after
convolution with the local LSF.  {\it Thick blue lines:}  
individual components (labelled with ticks), before convolution.  
{\it Solid black lines:} 
local continua.  Flux is in erg~cm$^{-2}$~s$^{-1}$~\AA$^{-1}$.
\label{fig-diowens}}
\end{center}
\end{figure}

{\it D~I:} 
With \Owens , the \DI\ lines for Ly7, 8, 9, 10 were 
fitted, and the \HI\ line parameters were also
permitted to vary in the fit.
For GH,
only the SiC2A data were used.  A fit from \Owens\ to Lyman 9 and 10 is shown
in Fig.~\ref{fig-diowens}.  
The reduced $\chi^2$ value is 1.04, with 1442 degrees of freedom.
CO included both SiC1B and SiC2A in the calculation, and had $\chi^2=1.65$ for 1878 degrees 
of freedom.  Results between the two measurements were consistent, showing that
including only the highest resolution data made no significant difference.

\begin{figure}
\begin{center}
\rotatebox{0}{
\epsscale{0.85}
\plotone{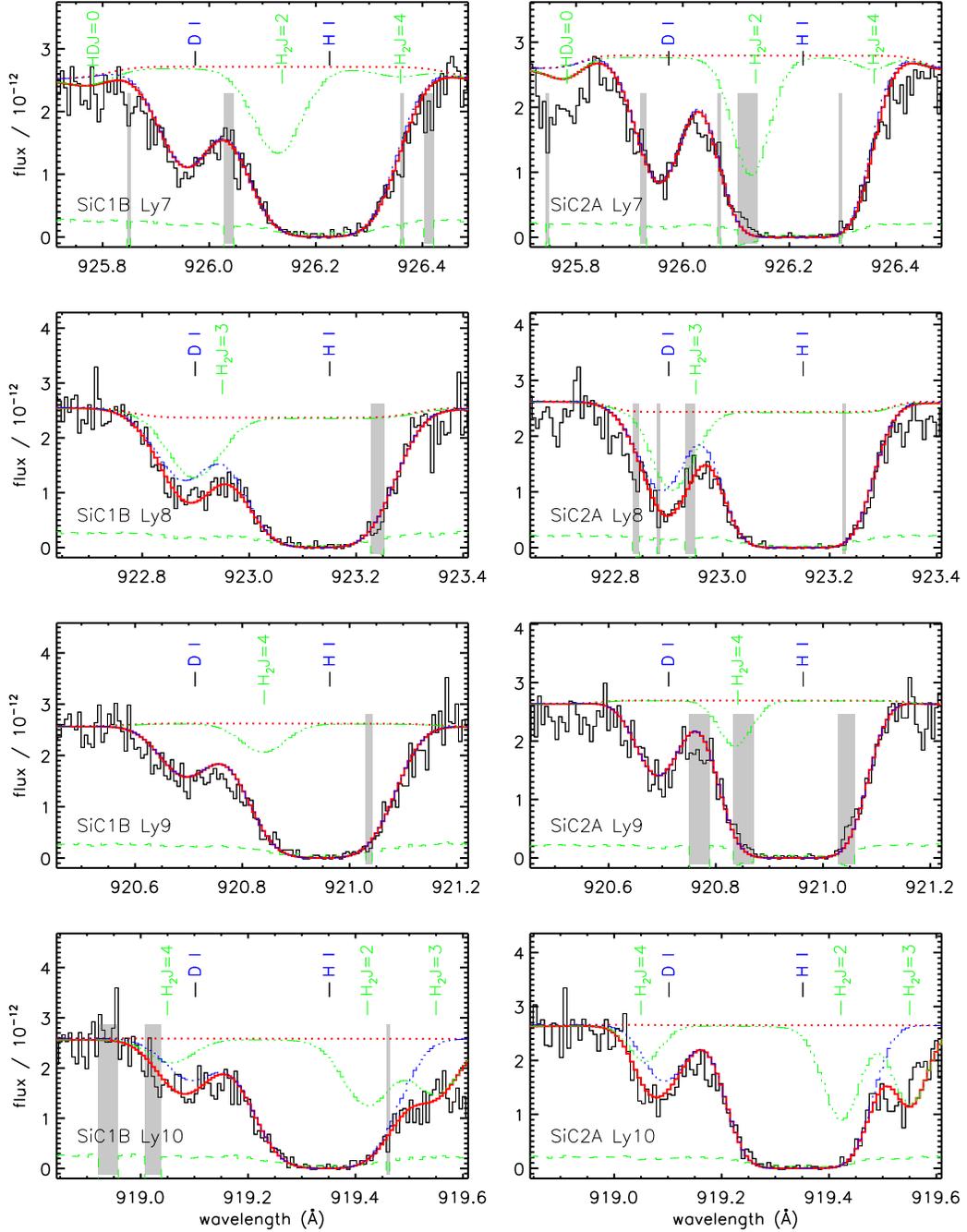}
}
\caption[\vpfit\ fit to higher order 
\HI , \DI\ and \HH\ profiles for the PG\,0038$+$199 line of sight]{
{\it Solid (red) line}: \vpfit\ combined fit to \hi , \di\ and \HH\  
(all profiles shown after convolution
with local LSF). 
{\it Dashed (green) line}: $1\sigma$ error array.
{\it Dot-dashed (green) line}: \HH\ profiles.
{\it Triple-dot-dashed (blue) line}: HI, DI profiles.
{\it Dotted (red) line:} Stellar continuum.
{\it Grey zones}: regions which were excluded from the fit, including geo-coronal emission and noise
spikes. The thick dashed and solid lines represent the continuum and the fit, respectively. 
\HH , \hi\ and \di\ transitions are marked by ticks.  Flux is in units of $10^{-12}$ erg
cm$^{-2}$ s$^{-1}$ \AA $^{-1}$. \label{fig-DIHIvpfitpg0038}}
\end{center}
\end{figure}

With \vpfit , 
the reduced $\chi^2 $ value
is 1.03 for 3635 degrees of freedom over Ly$\beta$,7,8,9,10, shown in
Fig.~\ref{fig-DIHIvpfitpg0038}.
In case the saturated \Lyb\ \DI\ line skews
the statistics, we also made a fit with only Ly7,8,9,10.
Results were not changed significantly.

Comparing the \Owens\ and \vpfit\ results, 
we agree on a column density $\log N(\DI ) = 15.75$.  Errors are discussed below.
The Doppler parameter, assumed to be a composite from several unresolved components
as for \HI , is $b=11.9\pm 1.6$~\kms\ ($2\sigma$ errors).

We checked the \DI\ column density using the AOD and COG methods.
For the AOD method we find $\log N(\DI ) = 15.68\pm 0.15$ ($1\sigma$ error in the mean).
A COG study for the SiC2A data (which have higher resolution than SiC1B) 
yields $\log N(\DI ) = 15.80^{+0.07}_{-0.06}$, $b=10.7^{+2.8}_{-1.8}$~\kms .
These values are consistent with our PF results.

\subsubsection{\HI\ and \DI\ Column density errors}

\citet{Hebrard02} identified and discussed a number of systematic effects which can affect
column density estimates.  Following their example,
we considered the potential systematic effects from the following sources:\\

{\it Continuum uncertainty:} 
Both \vpfit\ and \Owens\ can explicitly include local normalization corrections in their calculations of
statistical uncertainties.  

For the \Owens\ analyses, the local continuum is free to vary for each spectral
window, and is approximated by a polynomial of order $\sim 2-4$.  Uncertainties
from the continuum approximation are reflected in the $\chi^2$ statistics \citep[e.g.][]{Hebrard02,Oliveira03}.

For the \vpfit\ analysis, we allowed the continuum for each stellar model to have a zeroth order
correction factor 
in each of the 12 fitting regions (4 for \Lyb\ and 2 each for Ly7, 8, 9, 10), with a mean
value of $0.963\pm 0.041$ ($1\sigma$).  The correction factor is thus (just) consistent with
unity, and is
consistent with the continuum uncertainty from HD (\S~\ref{subsec-H2profilefit}).  We note that
the multiplicative corrections for the four \Lyb\ regions are $0.974\pm 0.004$,  $0.930\pm 0.005$,
$0.936\pm 0.006$ and $0.936\pm 0.007$ for the LiF1A, LiF2B, SiC1A and SiC2B segments respectively,
which may indicate a local systematic overestimation of the stellar model flux.  Continuum differences
between the models throughout the data except in the \Lyb\ region
are $\simlt 4-5$\% (\S~\ref{sec-stellarmodelpg0038}), and are
thus consistent with the local continuum variations in the \Lyb\ region for our profile fits. 
Continuum differences in the \Lyb\ region are discussed below.
The
stellar model including [He/H]=5 and metals produced smaller continuum normalization corrections 
at \HI~$\lambda 923$ and \DI~$\lambda 922$,
because it included contributions from \ion{N}{4}.  
However, the model with metals did not produce a statistically acceptable overall fit in the \Lyb\ region,
in contrast to the metal-free models.  In any case, the inclusion of metals in the stellar
model had no significant effect on either \HI\ or \DI\
column densities.
\\
{\it Stellar model:} We compare results for the
grid of stellar models  described in 
\S~\ref{sec-stellarmodelpg0038}.
The maximum deviation between the best fit and other 
models is $\pm 10$\% .  This occurs for the He-rich, metal-free models 
over $1025.2-1025.4$~\AA\ (\ion{He}{2}~$\lambda 1025$), or for the model
with [He/H]=5 and metals at \Lyb .
Typical deviations in the damping wings are on
the order of $\leq 5$\%.  
For the \Owens\ analysis, the \Lyb\ region was fitted 
using the stellar model grid without allowing for continuum corrections, producing
a systematic error of $\sim \pm 0.05$~dex ($2\sigma$).
For the \vpfit\ analysis, a similar calculation
employing local continuum zeroth order corrections
resulted in
variations in column density of $\Delta \log N(\HI ) = 0.02$ and $\Delta \log N(\DI ) = 0.002$ ($2\sigma$).
\\
{\it Line spread function:} We varied the LSF by 2~\kms\ ($\sim 1\sigma$), which produced changes
of $\Delta \log N(\HI ) = 0.005$ and $\Delta \log N(\DI ) < 0.001$.  
These differences are
both significantly 
smaller than the statistical errors (see below).  
We do not consider LSF uncertainties further in our source of error.\\ 
{\it \HH\ column densities:} We varied the \HH\ column
densities, trying the COG and PF results for $\log N(\HH )$ for the ensemble of $J=$0,1,2,3,4,5 and
HD  as rough $2\sigma$ upper and lower $\log N(\HH )$ bounds, respectively.  The resulting  column
density variations are $\Delta \log N(\HI ) = 0.03$ and $\Delta \log N(\DI ) = 0.01$ ($\sim 2\sigma
$).  \HH\ column density uncertainty is a major systematic error for our \HI\ and \DI\
column densities.

A $\chi^2$ grid in $\log N(\HI )$, $\log N(\DI )$ gives statistical errors of $\Delta \log N(\HI )
= 0.02$ and $\Delta \log N(\DI ) = 0.04$ (both $2\sigma$). Combining the systematic and statistical
errors in quadrature yields  $\log N(\HI ) = 20.41 \pm 0.08$ and $\log N(\DI ) = 15.75 \pm
0.08$ ($2\sigma$).

We are encouraged that the results are so similar between 
the \Owens\ and \vpfit\ analyses by
three different people, and including four independent \FUSE\ measurements.  
Furthermore, the results are robust, 
despite the use of binned {\it vs.} unbinned data,
the variation of continua, stellar models, LSF and \HH\ column densities.
This line of sight is only the third \DHrat\ target
for which only \fuse\ data are used to determine $N$(\HI ), the first two
being from \citet{Wood04}.

\subsubsection{Other estimates for \HI\ column density}
\label{sec-othernhiestimates}

The total H column density, $\log N(\HI + 2\HH )$,
defined here as $\Ht$, is
$20.48\pm 0.07$ ($2\sigma$), which
is in accord with that predicted using
\NaI\ \citep{Ferlet85}.  They employed 78 stellar sight lines to determine
$\log N(\NaI ) = 1.04 [\log N(\HI + \HH )] -9.09$ (their equation 1), 
with a correlation coefficient of 0.85 and
slope of $1.04\pm 0.08$.  Using
our measured column density of $\log N(\NaI )=12.19\pm 0.04$ ($2\sigma$)
yields $\log N(\Ht ) = 20.46^{+1.75}_{-1.42}$, with the uncertainty
dictated by the slope error in the \citeauthor{Ferlet85} relation.
The color excess $E(B-V)=0.037$ offers a prediction of 
$\log N(\Ht )=20.3$ with a scatter of $\sim 50$\% \citep{Savage79}.

\begin{figure}
\begin{center}
\rotatebox{0}{
\epsscale{1.05}
\plotone{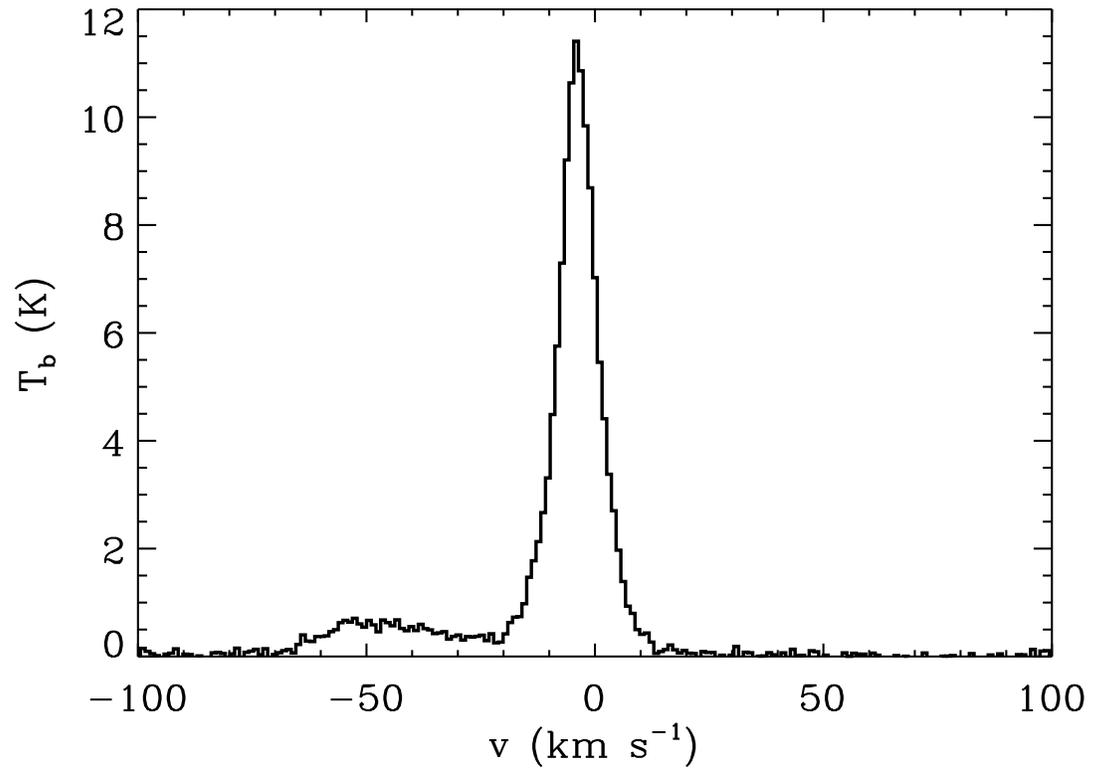}
}
\caption[\HI\ 21~cm flux for the PG\,0038$+$199 line of sight]{
The \HI\ 21~cm flux in units of brightness temperature {\it vs.} radial velocity toward \pgoo\ from the Leiden-Dwingeloo survey.
\label{fig-hi21cm}}
\end{center}
\end{figure}

We also examined
the \HI\ 21~cm profile from the Leiden-Dwingeloo \HI\ survey \citep{Hartmann97}, and
find $\log N(\HI ) =20.36$.  This result is for
a velocity resolution 1.03~\kms , 0.07~K temperature sensitivity and $0.5^\circ$ resolution on the sky
centered at Galactic coordinates $\ell=120^\circ$, $b=-42.5^\circ$ (13~arcmin from \pgoo ).
See Fig.~\ref{fig-hi21cm} for a velocity profile of 
the \HI\ 21~cm flux.  
We
integrated at the position of \pgoo\
between $-20<v<20$~\kms\ to avoid a secondary peak of emission on the blue side.
If we integrate over $-100<v<100$~\kms ,
$\log N(\HI ) =20.47$.  The beam size is likely large compared to variations in \HI , and
there is no way to tell from our \FUSE\ data how much \HI\ lies behind \pgoo , which would
increase the column density compared to what we could see in absorption with \FUSE .
Gas which is far behind \pgoo\ would likely be very far from the disk, given
its Galactic latitude.
Despite these uncertainties, the closeness of the
Leiden-Dwingeloo survey column density to that of our absorption measurement
lends supporting evidence to our result, and implies that the \HI\ is not very patchy in this
part of the sky.  In any case, there is no evidence for a column density
as large as $\log N(\HI )=20.57$, which would be necessary to bring the \DHrat\
ratio (discussed in the next section)
down to the Local Bubble value of $1.5\times 10^{-5}$.

\section{Results and Discussion}
\label{sec-discussion}

\begin{figure}
\begin{center}
\epsscale{1.00}
\plotone{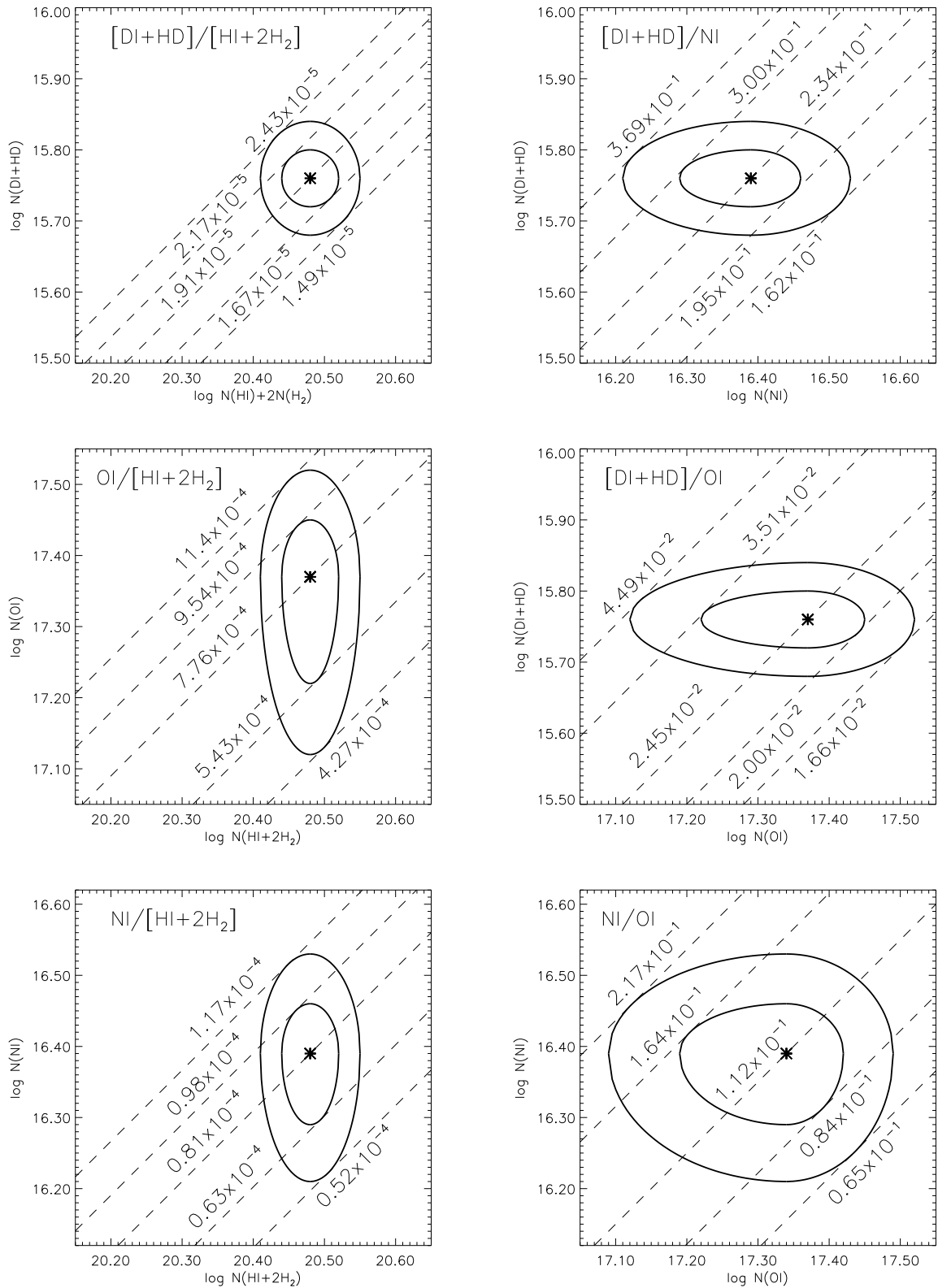}
\caption[Ratios of \DI +HD, \HI +2\HH, \OI , \NI\ column densities]{
Ratios, with $1\sigma$ and $2\sigma$ errors, for
[\DI +HD], [\HI +2\HH], \OI , \NI\ column densities.  Best fit values are shown by 
asterisks.
\label{fig-ratios}}
\end{center}
\end{figure}

Combining the results of our column densities for \NI\ and \OI\ with those
for  \di , \hi , HD and \HH , 
we obtain the relative column density ratios as listed in
Table~\ref{tab-ratios} and shown in Fig.~\ref{fig-ratios}.   We consider the ratios
between the column densities for \DI , \HI , \OI , \NI\ each in turn, 
beginning with ratios relative to hydrogen, and compare our results to those
in the literature. If we include \HH\ and HD contributions, we find $\log
[N(\Ht )] = 20.48\pm 0.07$ ($2\sigma $), and $\log [N(\DI) + N( {\mathrm {HD}}
)] = 15.76\pm 0.08$ ($2\sigma$). 
We therefore  quote various ratios 
including \HH\ and HD, as well as the traditional form \DHrat , plus 
the HD/\HH\ ratio.
We approximate
$1\sigma$ errors for HD and \HH\ as half the $2\sigma$ errors for contributions
to \DHrattot , \OHtot\ and \NHtot .  We
neglect HD for the total H contribution since (HD/[\HI +2\HH ]~$ <10^{-6}$).

\subsection{Column density ratios}
\label{subsec-columndensityratios}

\begin{figure}
\begin{center}
\rotatebox{0}{
\epsscale{1.05}
\plotone{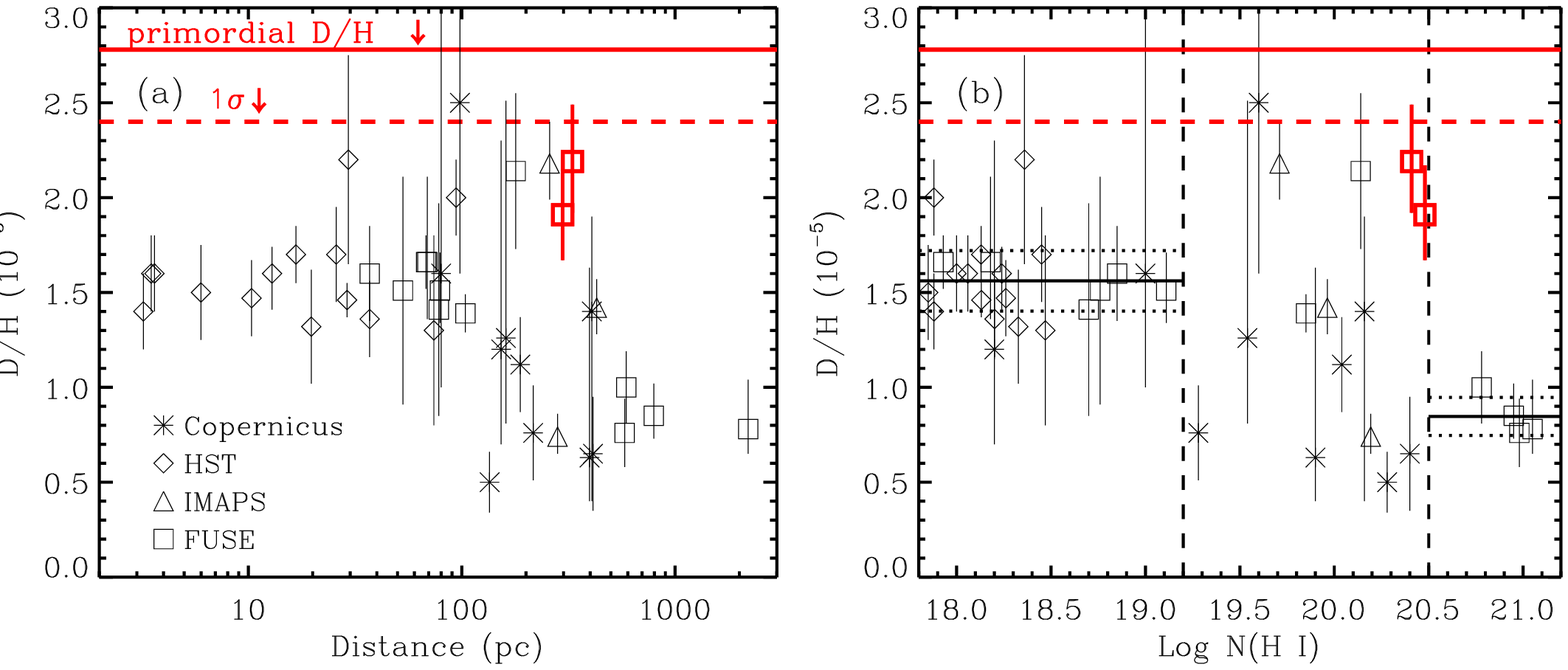}
}
\caption[]{
(a) \DHrat\ and $1\sigma$ error bars plotted {\it vs.} line-of-sight distance, using \DHrat\
measurements in \citet{Wood04} and this work (red).  Different symbols
are used for different sources of the \DI\ measurement. 
{\it Red open boxes}: \pgoo ,  representing \DHrat\ (upper box) and
\DHrattot\ (lower box), slightly offset in distance for clarity.
(b) \DHrat\ plotted
{\it vs.} line-of-sight \HI\ column density.  The symbols are the same
as in (a), specifically that  the {\it red open boxes} represent 
\pgoo , showing the \HI\ column
density (upper box) and the [\HI + 2\HH ] column density (lower box).
{\it Black horizontal and dotted lines:}
weighted means and $1\sigma $ standard deviations as a function of
$N(\HI )$ from \citeauthor{Wood04} 
{\it Red solid and dashed lines}: 
the primordial \DHrat\
value \citep{Kirkman03,Omeara01,Pettini01,Levshakov02} and lower $1\sigma $
standard deviation, respectively.
For intermediate values $19.2 < \log N(\HI ) < 20.5$, \DHrat\ appears variable.
\label{fig-dhplot}}
\end{center}
\end{figure}

\begin{figure}
\begin{center}
\rotatebox{0}{
\epsscale{1.0}
\plotone{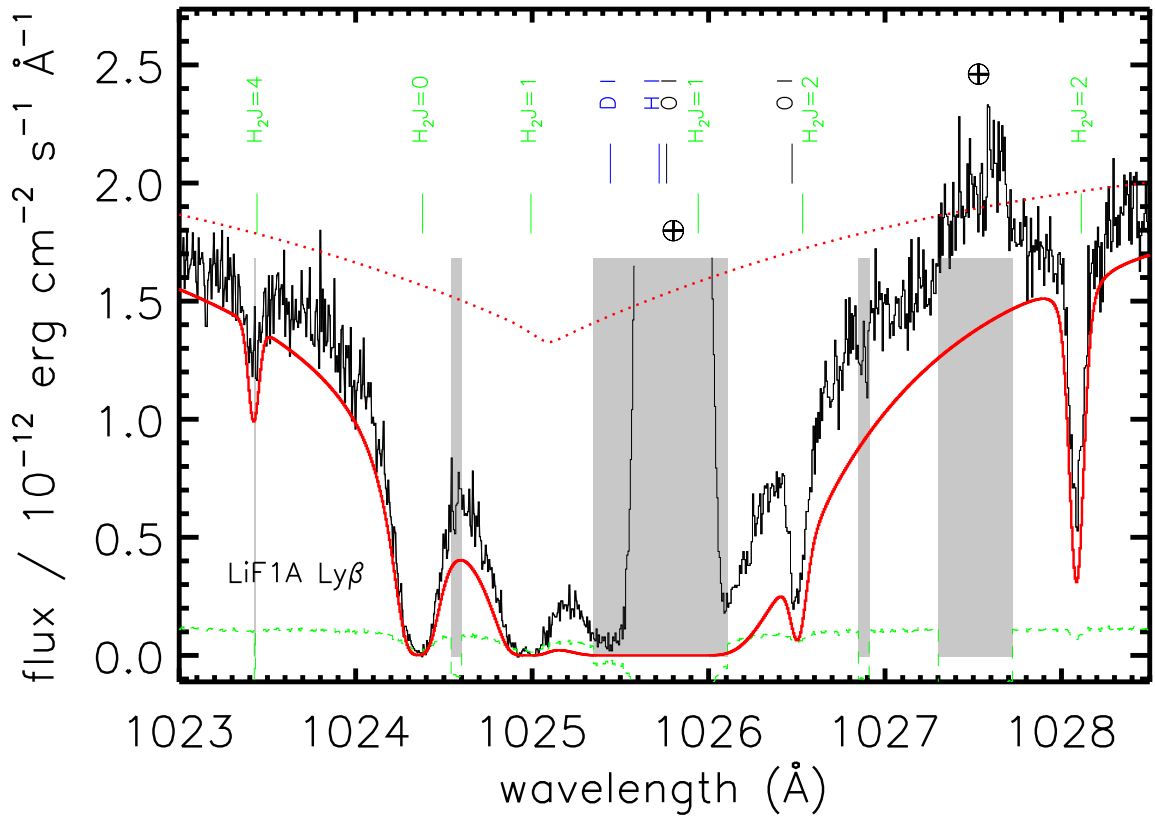}
}
\caption[Fit for $\log N(\HI )=20.83$]{
A model curve for $\log N(\HI )=20.83$ plotted with the LiF1A data, which would be
necessary to bring \DHrat =$0.85\times 10^{-5}$, in accord with
other high \HI\ column density sight lines.  Such a high column density is
firmly excluded by the data.
Line styles and symbols are as in Fig.~\ref{fig-HIvpfitpg0038}.
Column densities for
\DI , \OI\ and \HH\ have been kept constant at their best-fit values.
\label{fig-lybhi_nhi}}
\end{center}
\end{figure}

\DHrattot :  We obtain a value of
$(1.91^{+0.52}_{-0.42})\times 10^{-5}$ ($2\sigma$ errors), shown in Fig.~\ref{fig-dhplot},
which is 
significantly high compared to 
$\DHrat = (0.85\pm 0.09) \times 10^{-5}$ for
sight lines with
$\log N(\HI ) \geq 20.5$ and distances $d>500$~pc 
\citep{Wood04}.
The possibility of a low \DHrat\ value toward distant sight lines was first proposed by
\citet{Hebrard03}, based on \DO\ and \DN\ measurements, and previously published
\DHrat\ values.
\citeauthor{Wood04} corroborated the argument, compiling a large 
set of \DHrat\ measurements from the literature 
(though only four were for $\log N(\HI )>20.5$).  
If we exclude the molecular component, 
$\DHrat = (2.19^{+0.65}_{-0.50})\times 10^{-5}$ ($2\sigma $ errors)
making an even greater discrepancy with similar \HI\ column density sight lines.
We believe that our
\DHrattot\ and \DHrat\ ratios for \pgoo\ are 
robust, based on measurements from three individuals using two different
software packages and allowing for a variety of systematic errors.  Our
largest source of uncertainty is systematic errors in the \HI\ column density, because we do not
have access to \Lya .  
We would require 
$\log N(\HI +\HH ) = 20.83$ 
to have $\DHrattot = 0.85$, which is many $\sigma$ from our result.  Such a high value makes
a
poor fit to the data, and is shown in Fig.~\ref{fig-lybhi_nhi}.

\begin{figure}
\begin{center}
\rotatebox{0}{
\epsscale{1.0}
\plotone{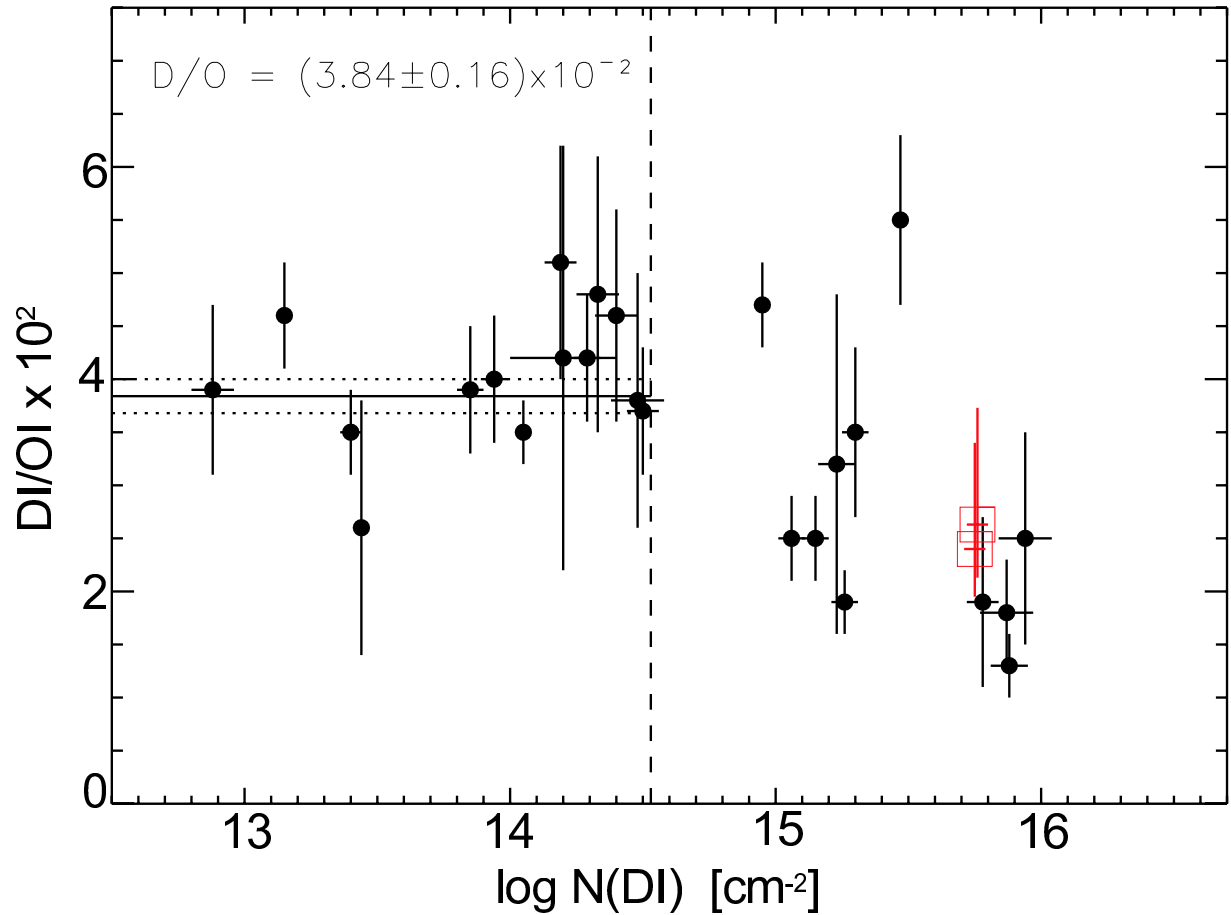}
}
\caption[\DO , \DN {\it vs.} $\log N(\DI )$]{DO\ {\it vs.} $\log
N(\DI )$ for a number of sight lines, with $1\sigma $ error bars, as in
\citet{Hebrard03} and including values from \citet{Wood04}. 
Values for \pgoo\ (both including and excluding 
HD contributions) are indicated by open boxes. 
The dotted line indicates
the limit inside the Local Bubble, inside which the \DO\ ratio is homogeneous.
\label{fig-doplot}}
\end{center}
\end{figure}

{\it \DOtot }:  Our result is $(2.63^{+2.18}_{-0.85})\times 10^{-2}$
($2\sigma $).
If  the HD 
contribution is excluded, the ratio drops to:  
$\DO = (2.40^{+2.19}_{-0.78})\times 10^{-2}$ ($2\sigma $).
For the sample $\log N(\DI )>15.5$ and distance
$d>300$~pc, our values are within the \citeauthor{Hebrard03} range for $(1.5\simlt
\DO  \simlt 2.5)\times 10^{-2}$, shown in Fig.~\ref{fig-doplot}, though we note that there are
only three values with that \DI\ column density threshold in their sample. The
$N(\OI )$ error dominates the uncertainty.

\OHtot :
We regard our \OI\ column density measurement with extreme caution, because it depends
entirely on one line at 974~\AA\ strongly blended with two \HH\ lines, dominated
by the SiC2A data. We 
compare our results with 
the \OH\ determination of \citet{Meyer98}, whose sample resembles our sight
line in \HI\ column  density.\footnote{The samples of \citet{Andre03} and 
\citet{Cartledge04} have significantly higher characteristic \HI\ and \HH\ column
densities than \pgoo .} Our value of  $\OHtot=(7.76^{+3.38}_{-3.49})\times
10^{-4}$ ($2\sigma$ errors) is  $>2\sigma $ higher than the local ISM value of 
$\OH=(3.43\pm 0.15)\times 10^{-4}$ 
($1\sigma$) from Meyer et al.
using the updated \OI\ oscillator strength of \citet{Welty99}.
If a \HI\ column density error were the sole cause of the O/H discrepancy, 
we would require $\log N(\HI )=20.82$ to bring \OH\ to $3.58\times 10^{-4}$,
or within the $1\sigma$
error of the revised \citeauthor{Meyer98} value.  However, such a high value for
$N(\HI )$ is highly unlikely, given our determinations from the \FUSE\ data and in light
of other $N(\HI )$ estimation methods discussed in \S~\ref{sec-othernhiestimates}.

Another source of \OH\ uncertainty
depends on the
accuracy of the \OI\ 974~\AA\ line oscillator strength, which is theoretically calculated and
is estimated to be good to 25\%
\citep{Biemont92,Wiese96}.  In addition to \pgoo ,
other objects in the literature for which $N(\OI )$ depends only on \OI~974 are
JL9, LSS 1274 \citep{Hebrard03,Wood04} and HD~90087 \citep{Hebrard04}.  
\footnote{HD~195965 \citep{Hoopes03} also has high \OH , but the \HI\ and \OI\ column densities 
have two independent measurements each from \HST , \FUSE\ and \IUE , and the star's position in the
Cygnus OB7 association allows for the possibility of containment within a metal-enriched cloud.}
They, too, exhibit
high values of \OH , though the local ISM value lies close to or within their $2\sigma$ formal errors.
If the \OI~974 oscillator strength were to increase by 15\% (0.06 dex), then the nominal
\OI\ column density toward \pgoo\ 
would drop to $\log N(\OI )=17.31$, with \OHtot~$=6.8\times 10^{-4}$,
putting the local ISM value within the $2\sigma$ error for \pgoo .
The nominal \DO\ value would increase from  to $2.40\times 10^{-2}$ to 
$2.57\times 10^{-2}$, which still lies
within the scatter of high column density data points from \citet{Hebrard03} (see above).
The \OI~974 oscillator strength appears consistent with that of \OI~1356 
in the case of HD~195965
\citep{Hoopes03,Hebrard04}, with the \citet{Hoopes03} uncertainty in $\log N(\OI )$ being
$+0.04,-0.06$ dex -- close to the value we require if the discrepant \OH\ value toward
\pgoo\ arises because of an oscillator strength difference.
Our \DHrat\ and \DO\ results are consistent with 
a larger variation in \DHrat\ than in \DO\ \citep{Hebrard03}.
However, 
we find \OH\ toward \pgoo\ anomalous, because the variations in \OH\ found by Meyer et al. were small,
and we cannot identify a specific reason why the measurement would be in error.
If we exclude the \HH\ contribution, the
\OH\ discrepancy is even larger: $\OH =(8.51^{+4.09}_{-3.86})\times 10^{-4}$
($2\sigma $).   Observations of additional lines of sight 
will show whether the combination of
\DI , \HI\ and \OI\ column densities found toward \pgoo\ produce an outlier,
are consistent with other sight lines or are subject to an unidentified error.

\begin{figure}
\begin{center}
\rotatebox{0}{
\epsscale{1.0}
\plotone{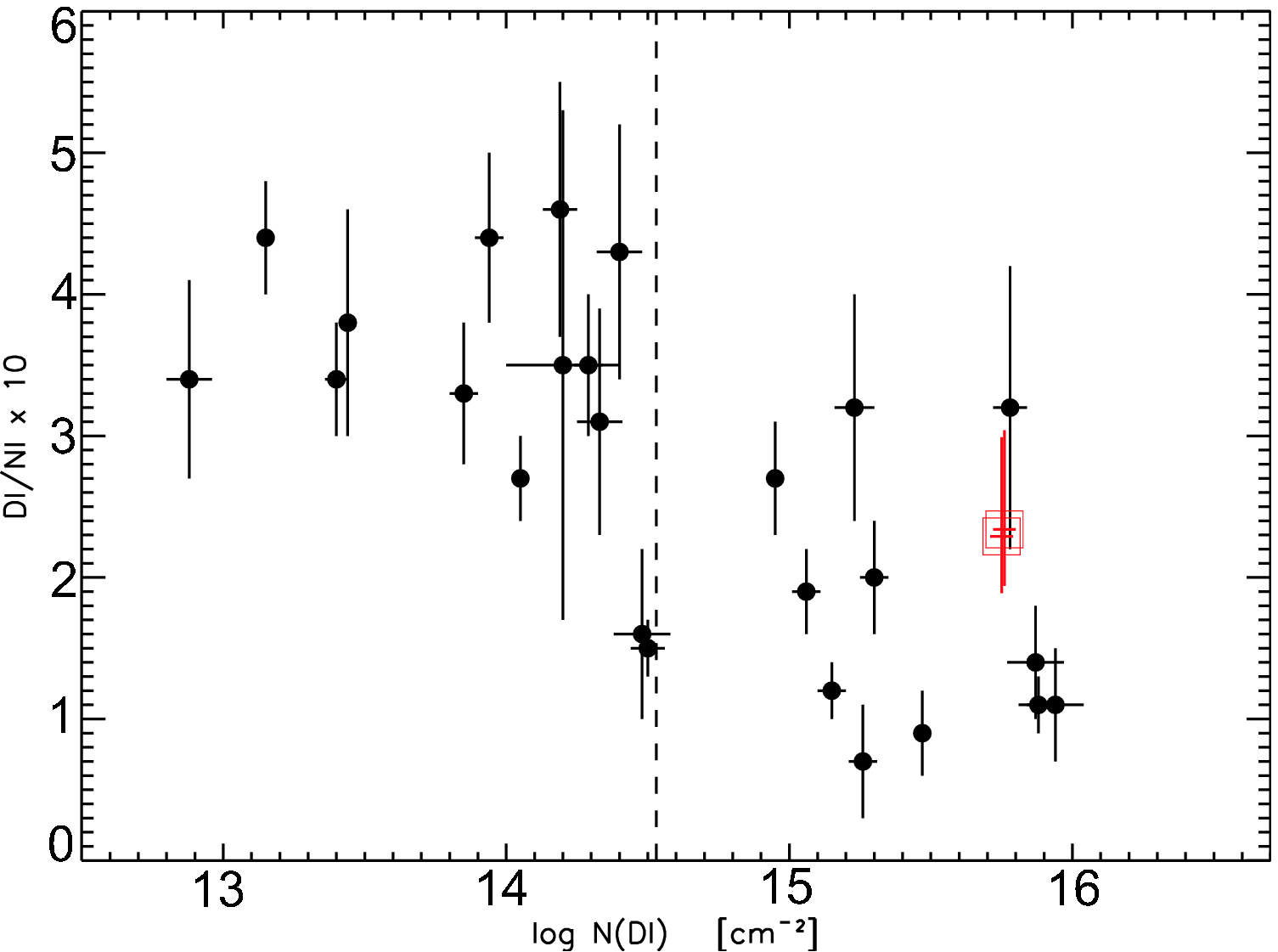}
}
\caption[\DO , \DN {\it vs.} $\log N(\DI )$]{
As in Fig.~\ref{fig-doplot}, but \DN\ {\it vs.} $\log N(\DI )$.  
\label{fig-dnplot}}
\end{center}
\end{figure}

{\it \DN }:
We find $\DNtot = (2.34^{+1.35}_{-0.72})\times 10^{-1}$ ($2\sigma$), which is 
high compared to the \citet{Hebrard03} values of $(1.0\simlt \DN \simlt 1.5)\times 10^{-1}$,
shown in
Fig.~\ref{fig-dnplot}, for $\log N(\DI ) \geq 15.5$.
Omitting the HD fraction results in
$\DN = (2.29^{+1.32}_{-0.71})\times 10^{-1}$ ($2\sigma$).

{\it \NHtot }:  We find values of
$(0.81^{+0.36}_{-0.29})
\times 10^{-4}$ and $\NH =(0.95^{+0.43}_{-0.34})
\times 10^{-4}$  (all $2\sigma $), which
are consistent within $1\sigma $ errors
of the ISM value of $\NH =(0.75\pm 0.04)\times 10^{-4}$ \citep{Meyer97}.

{\it \NO }:  Our value of $\NO = (1.12^{+1.05}_{-0.47})\times 10^{-1}$ ($2\sigma$) is
low compared to the ratio $(2.35\pm 0.15)\times 10^{-1}$ for \NH\ and \OH\ in the ISM
from \citet{Meyer97} and \citet{Meyer98}, due to the high \OH\ ratio  we find.
See the discussion above about the reliability of the \OI\ measurement.

HD/2H$_2$: The deuterium fraction in molecular form is $(2.08^{+0.34}_{-0.25})\times
10^{-6}$ ($2\sigma$ errors), and the column density ratio is HD/\HH~$=
(4.17^{+0.68}_{-0.49})\times 10^{-6}$ ($2\sigma$).   The HD and \HH\ column densities
place it within  the range of measurements at similar $N(\HH )$ from \citet{Spitzer74}
and \citet{Savage77}, a factor of $\sim 3$ above the expected ratio from charge
exchange reactions between H and D \citep{Liszt03}, but consistent with the scatter in values.

\subsection{Comparison with other high D/H lines of sight}
\label{sec-comparisonotherlos}

There are only two other noteworthy 
detections of $\DHrat \geq 2.1\times 10^{-5}$: $\gamma
^2$~Vel \citep{Sonneborn00} and Feige~110 
\citep{Friedman02}.\footnote{$\alpha$~Cru 
has high \DHrat\ but very large errors: $2.5^{+0.7}_{-0.9}\times 10^{-5}$ (presumed $1\sigma$ errors).
It has no reported \OI\ or \NI\ measurements, 
and has by far the lowest \HI\ column density among the high \DHrat\ sight lines, with 
$\log N(\HI )=19.6\pm 0.1$
\citep[presumed $1\sigma$,][]{York76}.} 
$\gamma ^2$~Vel exhibits both high and low ionization absorbers in a
7-member complex (3 \ion{H}{2}, 4 \HI ).  Its \HI\ absorption spans 
a velocity interval of $\Delta v =
32$~\kms , with a very low \HH\ fraction of  $1.4\times 10^{-5}$, and $\log N(\HI ) =
19.71\pm 0.03$ ($1\sigma$ errors),  $N(\DI ) = (1.12^{+0.15}_{-0.12})\times
10^{15}$ (90\% confidence limits)  and $\DN = (2.7\pm 0.4)\times 10^{-1}$  at 90\%
confidence or $1.65\sigma $ significance \citep{Cha00,Fitzpatrick94}. Although the \DN\
ratios are consistent within the errors for  $\gamma ^2$~Vel and \pgoo , in contrast,
\pgoo\ does not  exhibit numerous high ionization species (though it does show
\ion{S}{3}),  and has a much higher \HH\ fraction and \HI\ column density.  We cannot
compare \DO\ because the \OI\ column density  is difficult to measure for
both $\gamma^2$~Vel and \pgoo .
Feige~110 has only a slightly smaller \HI\ column density  than \pgoo\
($\log N(\HI ) = 20.14^{+0.13}_{-0.20}$, $2\sigma$ errors), and $\DHrat = (2.14\pm
0.82)\times 10^{-5}$ ($2\sigma$).  It is possible that \pgoo\ is slightly higher out of the
Galactic plane, and is sampling richer, infalling gas, one of several
explanations which we consider
in the next section.

Another way to characterize the nature of the \pgoo\ sight line versus the
$\gamma^2$~Vel sight line is in terms of the iron abundance.  
\citet{Jenkins86b}
have examined the iron abundances determined from N(\FeII )
measured by {\it Copernicus} as a function of the average density \nsubhavg .  
\nsubhavg\
is defined as $N({\mathrm H})=N(\HI )+2N(\HH )$ divided by the distance to the star.  For
\pgoo , $\nsubhavg =0.3^{+0.2}_{-0.1}$~cm$^{-3}$ ($1\sigma$)
and the relative iron abundance assumed to be
$\log (N(\FeII )/N({\mathrm H}))=-6.06$.  This iron abundance is higher than average, but
consistent within the uncertainties with the mid-density range results
presented by Jenkins and coworkers.  On the other hand, $\gamma^2$~Vel
has an iron abundance more than twice as great as \pgoo\ 
\citep{Jenkins86b,Fitzpatrick94}
and \nsubhavg\ is much
lower, 0.06~cm$^{-3}$.  
In the model discussed by 
\citet{Spitzer85} and used by \citeauthor{Jenkins86b}
to analyze the Fe abundance, the two sight lines are
quite different.  The $\gamma^2$~Vel sight line samples only relatively warm
diffuse gas with a density of $\sim 0.1$~cm$^{-3}$, 
while the \pgoo\ sight line samples
both warm gas and cold higher density fluctuations with a density of 
$\sim 0.7$~cm$^{-3}$ (\citeauthor{Spitzer85}). \citeauthor{Jenkins86b}
assumed that the properties of the grains
vary between the two densities and showed by fitting their data set that the
relative iron abundance would decrease a factor of four between the extreme
cases of only hot gas and only cold gas.  However, in this case, although the
iron abundance changes by 
more than a factor of two, as expected, these two
sight lines do not show an analagous change in the abundance of deuterium.   The
values of \nsubhavg\ and the Fe abundances are quite different for \pgoo\
versus $\gamma^2$~Vel, indicating that the grain properties along the two
sight lines are likely to be different, but there is no evidence for a similar
variation in the deuterium abundance.

\subsection{Variations in D/H: possible causes}

A number of different potential causes for variations in \DHrat\ have been discussed in the
literature.  All causes that have been proposed could produce
inhomogeneities in \DHrat , provided that the ISM mixing time is
sufficiently large.
\citet{Wood04} suggested that \DHrat\ variations occur on a scale where
ISM mixing produces inhomogeneities on scales larger than the Local Bubble,
which has a well-established \DHrat\ ratio,
yet small enough so that the sight lines 
to the most distant targets investigated for \DHrat\ pass through enough
inhomogeneous zones for \DHrat\ variations to average out to their observed low value
of $\approx 8.5\times 10^{-6}$.  The strongest challenge to the \citeauthor{Wood04}
model would be a sight line
with $N(\HI )>20.5$ (or in the Local Bubble) with a high 
\DHrat\ value.  \pgoo\ is the highest
\HI\ column density sight line with a high \DHrat\ value, just
below the $\log N(\HI )=20.5$ upper limit for the meso-mixing scale of \citeauthor{Wood04}
We now briefly discuss some of the proposed
    causes of \DHrat\ variability below, but see \citet{Lemoine99}  and
    \citet{Draine04} for details.

{\it a. Inhomogeneous Galactic infall:}
Material which has undergone little astration could fall inhomogeneously onto the
Galactic disk and produce the observed variation in \DHrat\ distribution.  One example is
from \citet{Chiappini02}, which predicts a gradient in \DHrat\ as a function of
Galactic radius.    More data are needed to determine the validity of the model
\citep[e.g.][]{Hebrard03}.

{\it b. Astration variations:}  Variations in the
    amount of stellar processing experienced by different regions of the
    ISM could result in \DHrat\ variability.
An anticorrelation between \DHrat\ and \OH\ has been suggested, on the basis
that gas is enriched in O as D is destroyed in stars \citep[e.g.][]{Steigman03}.
However, the suggestion was
tentative because the \DHrat\ sample size at the time was noted to be limited.  
\pgoo , with
its high values of both \DHrat\ and (if confirmed) \OH , clearly would be an exception to such an
anticorrelation.  
We share Steigman's view that more sight lines need to be analyzed to explore the
topic further.

{\it c. Variable deuterium depletion:}
Variations
of \DHrat\ could be caused by preferential depletion of D onto dust grains
relative to H.  \citet{Draine04} noted that such a process could
account for the variation in \DHrat\ in the ISM 
while allowing for the relative homogeneity
of abundances for N and O.  \citet{Lipshtat04}
showed that it is possible for dust 
to enhance the production rates of HD and D$_2$ {\it vs.}
\HH .
Observational evidence comes from \citet{Peeters04}, who found infrared spectral
evidence of deuterated polycyclic aromatic
hydrocarbons (PAHs), and pointed out the large deuterium excess in
primitive carbonaceous meteorites and interplanetary dust particles 
\citep[e.g.][]{Messenger00}.
\citeauthor{Draine04} also suggested that 
high velocity
shocks could also work in the opposite sense to destroy grains and increase \DHrat .

Dust depletion processes occur
in cold regions of the ISM where \HH\ is the dominant reservoir of H.
Preferential depletion of D onto dust requires particularly low
temperatures \citep{Draine04}, so one might expect lines of sight with
low \HH\ temperatures to have low \DHrat\ ratios.
Typical ISM \HH\ temperatures have been measured to have
mean values of
$T_{01}\sim 70$~K and {\it rms} scatter of $\sim 15$~K for over 80 lines of sight
\citep{Savage77,Rachford02}.  \citet{Wood04} noted similar temperatures toward the
high \HI\ column density sight lines JL~9 and LSS~1274,
and  proposed that
lines of sight which were recently shocked should have relatively high \DHrat .
\HH\ temperatures could be increased in star formation regions, such as
near  $\alpha$~Cru and $\gamma^2$~Vel, the latter being near the Vela supernova
remnant.  
As gas cools, deuterium could be preferentially depleted onto
PAHs or grains \citep{Draine04,Peeters04,Lipshtat04}.
The \HH\ temperature of $T_{01}=136\pm 21$~K toward \pgoo\ is unusually 
high compared to the mean ISM
value, so this is consistent with a high \DHrat\ value in the depletion model.
Quantitatively, such
depletion could reduce \DHrat\ by $1\times 10^{-5}$, which is consistent within the errors with
the 
difference in \DHrat\ between the \pgoo\ sight line and 
sight lines with $\log N(\HI )>20.5$.

However, we find no evidence of supernova remnants near \pgoo\ \citep{Stephenson02},
nor any sign of nearby anomalous X-ray emission \citep{Snowden95} or H$\alpha$ emission
\citep{Haffner03}.  The \HH\  temperature of \pgoo\ could  be raised by
the faint remnants of a planetary nebula, though. Two other similarly hot DO stars
have  H$\alpha$ detections of huge, old, faint PN:  PG~0109+111 \citep[][a search which
found no old PN around \pgoo ]{Werner97} and PG~1034+001 \citep{Rauch04}.

If a correlation exists between \HH\ temperature and deuterium depletion, it would have
to explain sight lines with low \DHrat\ and high \HH\ rotation temperatures.
$\delta$~Ori and $\epsilon$~Ori \citep{Jenkins00} exhibit such high rotation
temperatures with absorption components having
$T_{03}\sim 190-850$~K.  (The temperatures for $T_{01}$ are similar to $T_{03}$ for
those sight lines.  Reasons for the similarity are discussed in
\citeauthor{Jenkins00}, though we recognize that $T_{01}$ is usually
assumed to be the best estimate of the actual thermal temperature.)  
The \HI\ column densities toward $\delta$~Ori and $\epsilon$~Ori
are relatively high ($\log N(\HI ) =
20.193\pm 0.025$, $\log N(\HI )=20.40\pm 0.08$, $1\sigma$ errors), yet the \DHrat\
ratios are low ($0.74^{+0.12}_{-0.09}$, $0.65\pm 0.3 \times 10^{-5}$ respectively, $1\sigma$ 
errors).  Their \HH\
fractions are low, however ($7\times 10^{-6}$, $1.3\times 10^{-4}$), so a different
physical mechanism from that toward \pgoo\ may be at work.  In addition, \citet{Spitzer74} found the
rotation temperatures of 
$\gamma^2$~Vel and $\zeta$~Pup to be $T_{03}=964\pm 119$~K
and $T_{04}=1140\pm 154$~K, (both $1\sigma$) respectively.  (\citeauthor{Spitzer74}
state that their excitation temperatures for $T_{03}$ and $T_{04}$ fit all the
observed values of $N(\HH )$ for the $J$ levels included in the calculation; we
estimate $T_{01}$ from their \HH\ column densities for $J=0$ and 1 and 
their equation 1 to be $\sim 300\pm 150$~K and $\sim 350\pm 100$~K
respectively, which are still high.)
\DHrat\ toward $\gamma^2$~Vel is high, as
discussed in the previous section, but 
toward $\zeta$~Pup is $(1.42^{+0.25}_{-0.23})\times 10^{-5}$
\citep[90\% confidence, ][]{Sonneborn00},
which, given our uncertainties, is marginally below the value for \pgoo .
A larger data set is necessary to  investigate the relationship of \HH\
temperature and \DHrat\ ratio in further detail.
Additionally, if this depletion scenario is correct, then \DHrat\ may be correlated
with the abundance of refractory elements such as Fe.  However, as discussed in
\S~\ref{sec-comparisonotherlos}, there is no convincing
evidence of such a relationship for \pgoo .

It is clear that variation in \DHrat\ is significant 
outside of the Local Bubble, and its cause remains an open question.
Our understanding of variations in the \DHrat\
will improve as theory evolves and further
measurements of \DHrat , \HH\ and other ISM species are made at large distances in the Galactic
disk and halo.

\section{Conclusions}
\label{sec-conclusions}

We have observed the He-rich white dwarf \pgoo\ with \FUSE\ and Keck+HIRES and found the
following.

1. Keck HIRES data hint at unresolved structure on the order of $\leq 8$~\kms\ based on \NaI .

2. We have compared column densities for a number of ISM species 
based on a variety of measurements using profile fits from \Owens\ and \vpfit , 
curve-of-growth and apparent optical depth 
determinations, and found
broadly consistent results between the various methods.

3. The \HI , \DI\ and \HH\ column densities are $\log N(\HI )= 20.41\pm 0.08$, 
$\log N(\DI )= 15.75\pm 0.08$, $\log N(\HH ) = 19.33\pm 0.04$ 
with a \HH\ fraction $f_{\HH } = 0.14\pm 0.02$
($2\sigma$
errors quoted throughout the conclusions, unless otherwise noted).  
The measurement of $N(\HI )$ for a \DHrat\
target is only the third
based solely on \FUSE\ data.  The \HH\ fraction is high enough that we include effects for
it and HD in our column density ratios.

4. $\DHrattot = (1.91^{+0.52}_{-0.42})\times 10^{-5}$ 
is unusually high for a Galactic value.  Variations in \DHrat\ outside the Local
Bubble are significant.
$\NI / (\HI + 2\HH ) = (0.81^{+0.36}_{-0.29})\times 10^{-5}$ is 
consistent with the mean ISM value
\citep{Meyer97}.
$\DNtot = (2.34^{+1.35}_{-0.72})\times 10^{-1}$ is high
for $\log N(\DI ) \geq 15.5$ \citep[][though their sample size is small]{Hebrard03}.
$\OHtot = (7.76^{+3.38}_{-3.49})\times 10^{-4}$ is unusually high, with
$\log N(\OI )$ depending on a single line measurement.  However,
$\DO = (2.40^{+2.19}_{-0.78})\times 10^{-2}$ is consistent with three other
distant sight lines \citep{Hebrard03}.  Further observations should determine
whether the high \OH\ value remains an anomaly, arises as a result of some uncertainty
(e.g. \OI~974 oscillator strength)
or can in fact be statistically expected.

5. The HD/\HH\ ratio is $(4.17^{+0.68}_{-0.49})\times 10^{-6}$,
consistent with the scatter of
previously measured values for $\log N(\HH ) \approx 19.3$ \citep{Liszt03}.

6. A number of mechanisms which produce variation in \DHrat\ remain plausible.
The \HH\ excitation temperature of $143\pm 5$~K ($1\sigma$) may indicate past star formation
in the vicinity of \pgoo , 
liberating D which may otherwise
be depleted onto grains, thereby potentially
explaining the high \DHrat\ value.  
However, there are low \DHrat\ sight lines with even higher excitation temperatures.
More sight lines outside the Local Bubble 
need to be observed to characterize the \DHrat\ distribution and physical processes
associated with it.

\acknowledgements

This work was supported by the \FUSE\ \, IDT; \FUSE\ \, 
is operated 
under NASA contract NAS 5-26555.  
We thank S. Friedman, J. Linsky, G. Steigman  and B. Wood for useful
conversations on \DHrat\ in the ISM, 
B. Otte for assistance accessing WHAM data and W. Blair, P. Sonnentrucker for
a careful reading of the manuscript, E. Jenkins and J. Linsky for many suggestions to
improve the paper, P. Feldman and D. Sahnow for discussions about geocoronal
emission, E. Williger, an IR
spectroscopist,  for useful conversations on molecular spectra, and the
University of Akron for guest computer support.  We also thank
the anonymous referee for many suggestions to clarify the presentation of
this paper.  This work has been
done using the profile 
fitting procedures \Owens , developed by M. Lemoine and the French
\FUSE\ team, and \vpfit , provided by R. Carswell with much helpful advice. 
We acknowledge use of the SIMBAD database, operated at CDS in Strasbourg,
France, and native Hawaiians for enabling observations from Mauna Kea.
\vspace{1cm}

\appendix
\section{\bf Details on the reduction of \boldmath{$FUSE$} data}
\label{appendix-reduction}

We present details on the reduction of \FUSE\ data, as the use of \vpfit\
creates a need for accurate $\sigma$ errors, a desire for the most accurate wavelength
calibration possible and requirement for knowledge of the LSF as a function of segment and
wavelength.

\subsection{Correction of $\sigma$ values}

\begin{figure}
\epsscale{1.05}
\plotone{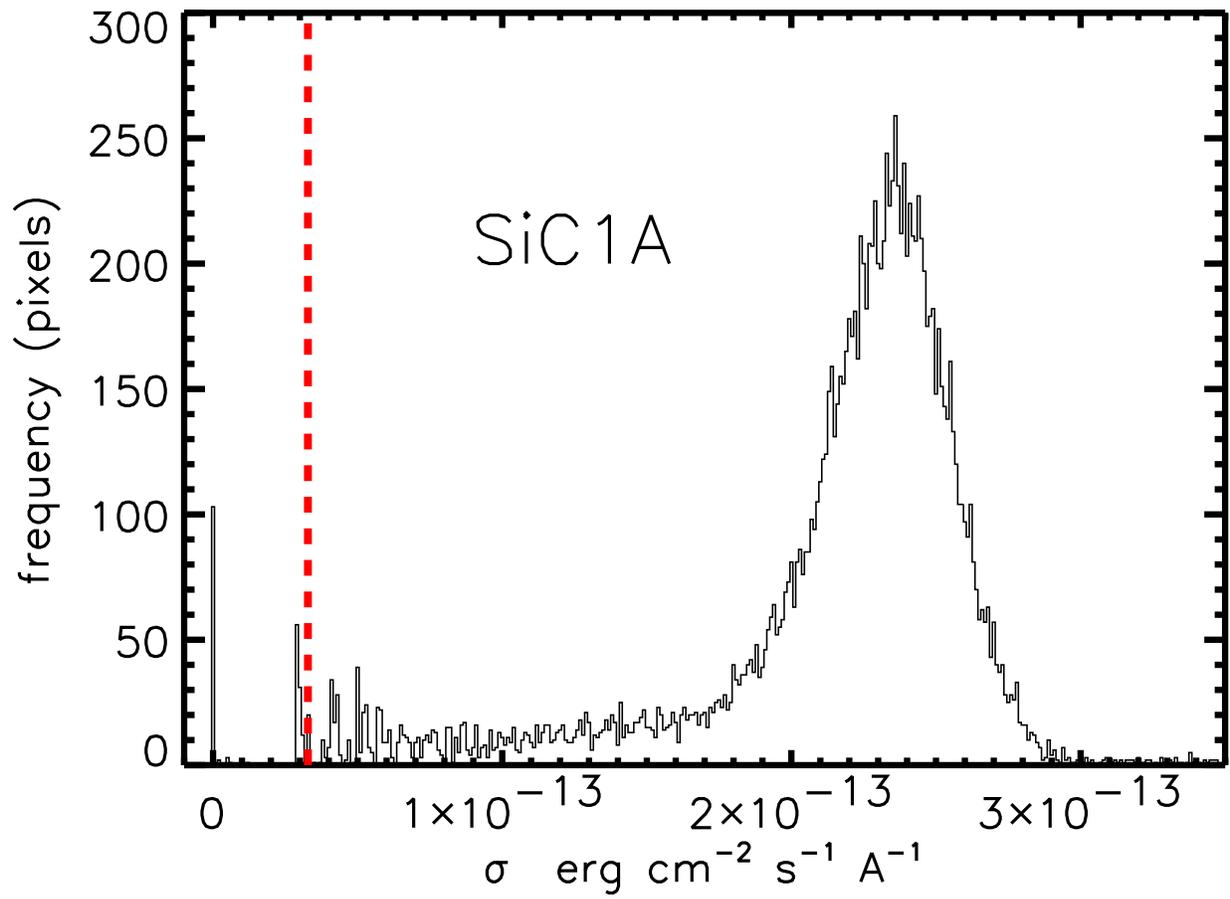}
\caption{Example distribution of $\sigma$ error values
for SiC1A segment.  The dotted line shows the point at which
we set a floor for $\sigma$ values for SiC1A data.
\label{fig-sigmas}
}
\end{figure}

We established a minimum value for the $\sigma$ array based on
the distribution of $\sigma$ values for each segment.  Individual 
pixels with values of $\sigma$ which
are unrealistically low make it difficult or impossible to obtain acceptable
statistical results for profile fitting.
The mean flux
and signal to noise ratio per pixel 
for the eight segments before modification of the
$\sigma$ values, for $\lambda > 915$~\AA\ and where $\sigma>0$, ranges over
$1.4-1.8\times 10^{-12}$~erg~cm$^{-2}$~s$^{-1}$~\AA$^{-1}$
and $5.0-18.2$.\footnote{Unbinned pixels, $\sim 0.0066$~\AA /pixel, are used throughout this analysis
unless otherwise noted.}  We plotted histograms of the unmodified $\sigma$ values,
and noted narrow, secondary 
peaks of low $\sigma$ values significantly (4--5 full width half maxima) below  
the main peak of the $\sigma$ distribution.  There were a number of instances
with $\sigma=0$.
We therefore set a minimum $\sigma$ value for each \FUSE\ detector segment.  The floors
in the $\sigma$ distribution
range from 
$1.4-5.5\times 10^{-14}$~erg~cm$^{-2}$~s$^{-1}$~\AA$^{-1}$, which affects pixel
fractions of 0.4-1.5\% per detector segment.  An example is shown in Fig.~\ref{fig-sigmas}.

We then calculated an empirical correction factor to the $1\sigma$ error arrays propagated
by the CalFUSE pipeline and affected by any rebinning of the data, fixed pattern noise etc.  
The correction is necessary to ensure proper $\chi^2$ probabilities
for the profile fitting, and is recommended in \vpfit\ documentation.
We first divided the data in each detector segment
by the continuum to normalize the spectrum.
We then binned the data for each segment, starting with increments of 5 pixels,
divided the root mean
square ($rms$) by the mean of the $1\sigma$ error array $\langle \sigma \rangle$
for all bins in a segment, and examined the distribution of $rms/\sigma$ ratios as a
function of bin size to find a characteristic ratio.
The process was repeated for bins of 6,7,8,...,100 pixels, in increments of one pixel.

The mode of the distribution is the most stable characteristic quantity of the distribution
of $rms/\sigma$ ratios, because
the mean and median can be skewed by bins containing large absorption features.
We therefore 
plotted the mode of $rms/\langle \sigma \rangle$ as a function of bin size 
to look for a trend in the values.  
A
resolution element is $\sim 10$ unbinned pixels, 
therefore the mode grows quickly with bin size up to $\sim 10-15$ pixels for
small bins, then stabilizes as the bin size increases to 20-40 pixels, which
is characteristic of the size of
spectral intervals between absorption lines. (There are deemed to be no  significant
intrinsic emission lines in the stellar spectrum.) 
At the bin size $\simgt 40$ pixels,
the ratio increases again because the wavelength range covered becomes characteristically
larger than the interval between absorption features.   
There was always a plateau in the value of the mode at $\sim 20-40$ pixels in 
bin width.
We took the value of the plateau to be
the $\sigma$ correction factor.
The correction factors (plateaus) are
1.4, 1.1, 1.3, 1.3, 1.1, 1.1, 1.2, 1.1 for LiF1A, LiF1B, LiF2A, LiF2B, SiC1A,
SiC1B, SiC2A, SiC2B respectively.  

The large correction factor for LiF1A is likely due to
fixed pattern noise.    Because focal plane splits (which
move the spectrum around on the detector) were not done for these observations,
fixed pattern noise should be present.  In the other segments, some movement of \pgoo\ in
the LWRS aperture is expected.  However, LiF1A is used for guiding, so the movement of
the spectrum between exposures is smaller relative to the other segments.  LiF1A data are therefore
predicted to suffer the most from
fixed pattern noise, hence the large $\sigma$ correction factor.

\subsection{Correction of velocity scale}

\begin{figure}
\epsscale{1.00}
\plotone{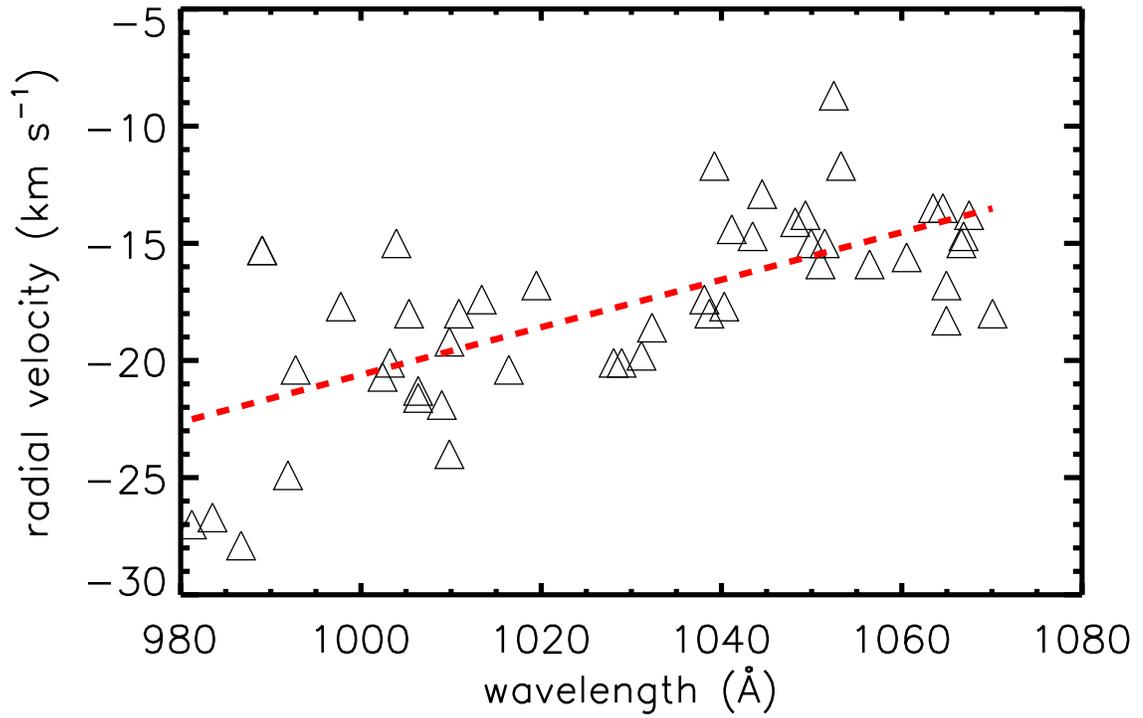}
\caption{Example distribution of \HH\ radial velocities for
segment LiF2B, which had the largest correction to the slope
of the wavelength solution (10~\kms ~100~\AA$^{-1}$).
The dashed line shows the best linear fit.  The slope was removed
and a linear offset applied to match the radial velocity derived
from our Keck observations of \NaI\ (-5.1~\kms ).
\label{fig-velcorrection}
}
\end{figure}

We made trial profile fits of \HH\ lines with \vpfit\  to
evaluate the accuracy of the relative wavelength scale within each
segment.   Radial velocity uncertainties for each \HH\ line are $\sim
1$~\kms\ (0.5 pixels). With $\sim 50$ \HH\ lines per segment (save for the
long wavelength ones LiF1B and LiF2A), linear dispersion corrections of
2--10~\kms~100~\AA$^{-1}$ were applied, 
plus a zero point offset to make the \HH\ velocities match the \NaI\
heliocentric velocity of -5.1~\kms\ measured from our Keck data. 
See Fig.~\ref{fig-velcorrection} for an example.
Standard deviations of
radial velocity residuals after applying the linear corrections to each
segment were 1--4~\kms\ (0.5--2 pixels).

\section{\bf Determination of the line spread function using \HH\ lines}
\label{appendix-h2lsf}

We made an independent determination of the 
\fuse\ instrumental profile (line spread function, LSF), which is poorly known.
The LSF
is suspected to
vary with detector/channel/side segment, wavelength, time, pointing accuracy,
and how
accurately individual exposures are cross-correlated and summed \citep{Kruk02}.
The problem was also studied  by \citet{Hebrard02}.
\citet{Wood02} attempted to solve for the LSF, coming up with a double Gaussian fit.
We attempted to incorporate their solution into the \fuse\ LSF for profile fits with
\vpfit , but found that a single Gaussian gave better results based on the \chisq\ statistic.
However, the FWHM of the Gaussian which gave the best fit varied between \fuse\ segments.
We therefore decided to derive the \fuse\ LSF by using \HH\ lines as fiducials.
This was done by assuming
the results from a combination of profile fitting with 
\Owens\ (which leaves the LSF as a free parameter) and the COG method, and varying
the width of
a Gaussian LSF in \kms\ 
for a each of a large number of \HH\ lines.  We only used unblended lines of the
\HH\ $J=2,3,4$
rotation levels, because the $J=0$ and 1 transitions are in the damping part of the
curve of growth and are thereby mostly resolved, making them useless for assessing
the LSF.  (This is borne out by their showing
uniformly inflated LSF FWHM values compared to
the lower column density $J=2,3,4$ levels.)  The number of \HH\ lines fitted per segment was 29--45 except for
the two long wavelength segments (LiF1B and LiF2A), which only had 4 and 7 respectively.

\begin{figure}
\begin{center}
\rotatebox{0}{
\epsscale{1.00}
\plotone{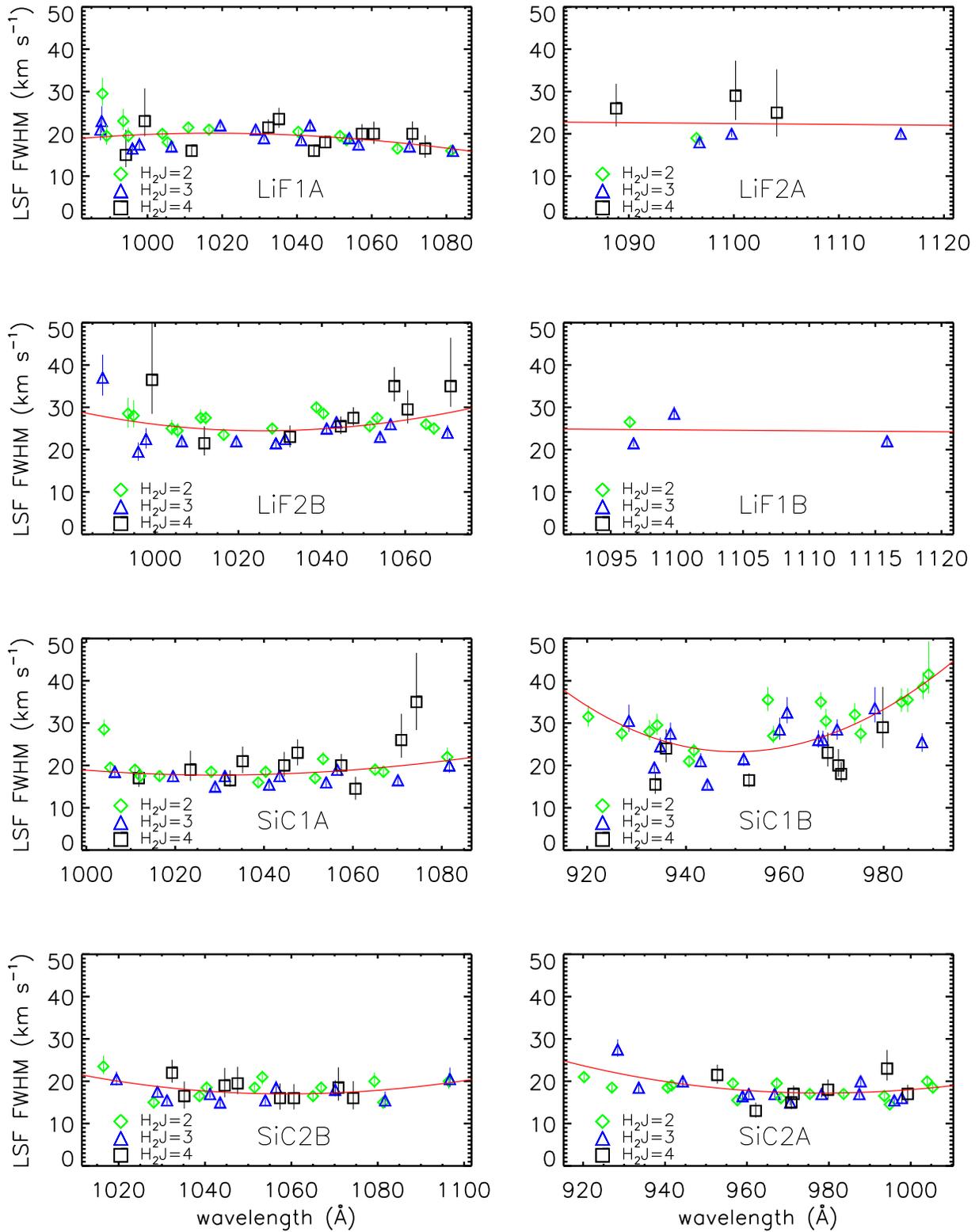}
}
\caption[]{Fit to LSF for all segments, based on \HH\ lines of rotation level $J=2,3,4$.
Error bars are $1\sigma$.
\label{fig-lsf}}
\end{center}
\end{figure}

\vpfit\ can accommodate a line spread function which varies as a function of wavelength, 
and uses a power series in
FWHM in \AA\ as a function of wavelength.  We therefore converted the FWHM from \kms\ to \AA\
and fitted a quadratic function to each of the segments using the interactive IRAF 
routine {\sc curfit}, 
except in the case of LiF1B and LiF2A,
which we fitted with simple constants in \kms .  
The root mean square of the LSF residuals ranged from
0.007--0.018~\AA\ ($\sim 2.0-5.6$~\kms ), with by far the worst fit for SiC1B.  Plots for the
LSF determinations are in Fig.~\ref{fig-lsf}.

\clearpage
\begin{deluxetable}{lcl}
\tablewidth{0pc}
\tablecaption{\pgoo\ stellar properties
\label{tab-stellarprops}}
\tablehead{ 
\colhead{Quantity} & \colhead{Value} & \colhead{Reference} }
\startdata
Galactic coordinates	   & $\ell = 119.79^\circ$, $b=-42.66^\circ$ & \citet{Green86}	  \\
Spectral type		   & DO                                      & \citet{Wesemael93} \\
V			   & $14.544\pm0.016$	                     & \citet{Williams01} \\
$T_{eff}$		   & 115000$\pm 11500$~K	                     & \citet{Dreizler96} \\
$\log g$		   & $7.5\pm0.3$                             & \citet{Dreizler96} \\
Mass			   & 0.59~M$_{\odot}$                        & \citet{Dreizler96} \\
log(H/He)		   & $-0.7$	                             & \citet{Dreizler97} \\
log(N/He)	           & $-3$:                                   & \citet{Dreizler96} \\
E(B-V)			   & $\simlt 0.037$	                     & \citet{Schlegel98} \\
Distance		   & $297^{+164}_{-104}$~pc	             & this work          \\
\enddata
\vspace{2mm}
\end{deluxetable}

\clearpage
\begin{deluxetable}{crcl}
\tablewidth{0pc}
\tablecaption{\HH\ and HD lines used for COG analysis.
\label{tab-h2hdlines}}
\tablehead{ 
\colhead{} & \colhead{}      
&\colhead{} &\colhead{Equiv Width}\\
\colhead{Species}    & \colhead{$\lambda$ (\AA )\tablenotemark{a}}     
&\colhead{$\log f\lambda$\tablenotemark{a}} &\colhead{(m\AA )} }
\startdata
H$_2$~$J$=0 &  923.990 & 0.751 & $125.0\pm  6.3$		  \\
H$_2$~$J$=0 &  931.069 & 0.990 & $137.7\pm 13.2$		  \\
H$_2$~$J$=0 &  962.985 & 1.101 & $200.6\pm  7.7$		  \\
H$_2$~$J$=0 &  981.442 & 1.306 & $265.0\pm 19.7$		  \\
H$_2$~$J$=0 & 1077.141 & 1.100 & $256.8\pm  8.2$		  \\
H$_2$~$J$=0 & 1108.139 & 0.267 & $116.5\pm  5.6$		  \\
H$_2$~$J$=1 &  924.649 & 0.580 & $106.5\pm  8.6$		  \\
H$_2$~$J$=1 &  925.180 & 0.256 & \phantom{1}$ 98.9\pm  6.9$ \\
H$_2$~$J$=1 &  932.273 & 0.344 & $117.3\pm  9.1$		  \\
H$_2$~$J$=1 &  982.839 & 0.824 & $222.0\pm  9.0$		  \\
H$_2$~$J$=1 &  992.812 & 0.883 & $277.8\pm 11.2$		  \\
H$_2$~$J$=1 & 1003.300 & 0.929 & $282.6\pm  9.3$		  \\
H$_2$~$J$=1 & 1108.644 & 0.078 & $155.9\pm  6.2$		  \\
H$_2$~$J$=2 &  920.248 & 0.187 & \phantom{1}$ 38.0\pm  4.2$ \\
H$_2$~$J$=2 &  927.026 & 0.329 & \phantom{1}$ 54.9\pm  4.8$ \\
H$_2$~$J$=2 &  932.614 & 0.647 & \phantom{1}$ 61.2\pm  7.8$ \\
H$_2$~$J$=2 &  933.245 & 0.803 & \phantom{1}$ 82.6\pm  6.9$ \\
H$_2$~$J$=2 &  934.151 & 0.317 & \phantom{1}$ 45.1\pm  4.1$ \\
H$_2$~$J$=2 &  940.632 & 0.749 & \phantom{1}$ 51.1\pm  3.0$ \\
H$_2$~$J$=2 &  941.606 & 0.498 & \phantom{1}$ 53.7\pm  3.6$ \\
H$_2$~$J$=2 &  956.586 & 0.941 & \phantom{1}$ 63.4\pm  4.6$ \\
H$_2$~$J$=2 &  957.660 & 0.661 & \phantom{1}$ 58.2\pm  4.6$ \\
H$_2$~$J$=2 &  965.794 & 1.495 & \phantom{1}$ 83.4\pm  6.7$ \\
H$_2$~$J$=2 &  967.283 & 1.530 & \phantom{1}$ 90.9\pm  5.4$ \\
H$_2$~$J$=2 &  968.297 & 0.844 & \phantom{1}$ 68.9\pm  4.9$ \\
H$_2$~$J$=2 &  975.351 & 0.810 & \phantom{1}$ 62.3\pm  4.1$ \\
H$_2$~$J$=2 &  983.594 & 1.054 & \phantom{1}$ 70.3\pm  3.8$ \\
H$_2$~$J$=2 &  984.867 & 0.905 & \phantom{1}$ 64.4\pm  4.6$ \\
H$_2$~$J$=2 &  987.978 & 1.557 & \phantom{1}$ 99.1\pm  5.4$ \\
H$_2$~$J$=2 &  989.092 & 0.903 & \phantom{1}$ 48.2\pm  4.9$ \\
H$_2$~$J$=2 &  993.552 & 1.228 & \phantom{1}$ 96.6\pm  5.9$ \\
H$_2$~$J$=2 &  994.877 & 0.935 & \phantom{1}$ 92.3\pm  6.0$ \\
H$_2$~$J$=2 & 1003.988 & 1.222 & \phantom{1}$ 81.0\pm  4.4$ \\
H$_2$~$J$=2 & 1005.397 & 0.998 & \phantom{1}$ 64.1\pm  3.8$ \\
H$_2$~$J$=2 & 1009.030 & 1.200 & \phantom{1}$ 64.0\pm  4.3$ \\
H$_2$~$J$=2 & 1010.135 & 1.147 & \phantom{1}$ 57.2\pm  5.1$ \\
H$_2$~$J$=2 & 1010.945 & 1.381 & \phantom{1}$ 94.6\pm  4.0$ \\
H$_2$~$J$=2 & 1016.466 & 1.016 & \phantom{1}$ 75.8\pm  3.7$ \\
H$_2$~$J$=2 & 1038.695 & 1.234 & \phantom{1}$ 97.6\pm  5.2$ \\
H$_2$~$J$=2 & 1040.372 & 1.030 & \phantom{1}$ 79.5\pm  4.2$ \\
H$_2$~$J$=2 & 1051.505 & 1.168 & \phantom{1}$ 77.8\pm  4.0$ \\
H$_2$~$J$=2 & 1053.291 & 0.980 & \phantom{1}$ 74.9\pm  3.7$ \\
H$_2$~$J$=2 & 1065.001 & 1.057 & \phantom{1}$ 75.8\pm  4.7$ \\
H$_2$~$J$=2 & 1066.907 & 0.881 & \phantom{1}$ 85.3\pm  4.3$ \\
H$_2$~$J$=2 & 1079.227 & 0.868 & \phantom{1}$ 67.9\pm  5.5$ \\
H$_2$~$J$=2 & 1081.268 & 0.709 & \phantom{1}$ 79.4\pm  3.7$ \\
H$_2$~$J$=3 &  928.443 & 0.481 & \phantom{1}$ 32.4\pm  4.8$ \\
H$_2$~$J$=3 &  933.588 & 1.260 & \phantom{1}$ 41.4\pm  3.7$ \\
H$_2$~$J$=3 &  934.800 & 0.820 & \phantom{1}$ 51.2\pm  4.8$ \\
H$_2$~$J$=3 &  942.970 & 0.729 & \phantom{1}$ 47.2\pm  3.9$ \\
H$_2$~$J$=3 &  951.681 & 1.079 & \phantom{1}$ 49.1\pm  5.1$ \\
H$_2$~$J$=3 &  958.953 & 0.930 & \phantom{1}$ 50.0\pm  3.9$ \\
H$_2$~$J$=3 &  960.458 & 0.674 & \phantom{1}$ 46.0\pm  3.9$ \\
H$_2$~$J$=3 &  966.787 & 0.879 & \phantom{1}$ 39.5\pm  5.0$ \\
H$_2$~$J$=3 &  967.677 & 1.340 & \phantom{1}$ 58.1\pm  4.6$ \\
H$_2$~$J$=3 &  970.565 & 0.974 & \phantom{1}$ 53.4\pm  4.5$ \\
H$_2$~$J$=3 &  978.223 & 0.818 & \phantom{1}$ 46.2\pm  3.8$ \\
H$_2$~$J$=3 &  987.450 & 1.406 & \phantom{1}$ 48.8\pm  4.3$ \\
H$_2$~$J$=3 &  987.772 & 0.945 & \phantom{1}$ 51.7\pm  4.9$ \\
H$_2$~$J$=3 &  995.974 & 1.218 & \phantom{1}$ 65.0\pm  4.3$ \\
H$_2$~$J$=3 &  997.830 & 0.942 & \phantom{1}$ 59.4\pm  4.9$ \\
H$_2$~$J$=3 & 1006.417 & 1.199 & \phantom{1}$ 62.4\pm  4.1$ \\
H$_2$~$J$=3 & 1028.991 & 1.250 & \phantom{1}$ 55.2\pm  3.4$ \\
H$_2$~$J$=3 & 1031.198 & 1.059 & \phantom{1}$ 50.5\pm  3.2$ \\
H$_2$~$J$=3 & 1041.163 & 1.216 & \phantom{1}$ 68.6\pm  3.9$ \\
H$_2$~$J$=3 & 1043.508 & 1.052 & \phantom{1}$ 61.1\pm  3.5$ \\
H$_2$~$J$=3 & 1053.982 & 1.150 & \phantom{1}$ 54.8\pm  3.2$ \\
H$_2$~$J$=3 & 1056.479 & 1.006 & \phantom{1}$ 57.4\pm  3.2$ \\
H$_2$~$J$=3 & 1067.485 & 1.028 & \phantom{1}$ 54.1\pm  3.2$ \\
H$_2$~$J$=3 & 1070.148 & 0.909 & \phantom{1}$ 56.1\pm  3.6$ \\
H$_2$~$J$=3 & 1099.795 & 0.448 & \phantom{1}$ 46.6\pm  3.2$ \\
H$_2$~$J$=3 & 1115.907 &-0.081 & \phantom{1}$ 36.9\pm  3.0$ \\
H$_2$~$J$=4 &  935.969 & 1.264 & \phantom{1}$ 29.5\pm  4.1$ \\
H$_2$~$J$=4 &  952.765 & 1.420 & \phantom{1}$ 33.7\pm  4.1$ \\
H$_2$~$J$=4 &  962.158 & 0.921 & \phantom{1}$ 19.4\pm  3.6$ \\
H$_2$~$J$=4 &  968.671 & 1.097 & \phantom{1}$ 26.9\pm  4.3$ \\
H$_2$~$J$=4 &  971.392 & 1.533 & \phantom{1}$ 33.1\pm  4.1$ \\
H$_2$~$J$=4 &  979.808 & 1.095 & \phantom{1}$ 22.8\pm  3.4$ \\
H$_2$~$J$=4 &  994.234 & 1.134 & \phantom{1}$ 21.3\pm  4.4$ \\
H$_2$~$J$=4 &  999.272 & 1.217 & \phantom{1}$ 23.5\pm  3.7$ \\
H$_2$~$J$=4 & 1011.818 & 1.154 & \phantom{1}$ 35.8\pm  3.0$ \\
H$_2$~$J$=4 & 1023.441 & 1.031 & \phantom{1}$ 22.2\pm  3.6$ \\
H$_2$~$J$=4 & 1032.355 & 1.247 & \phantom{1}$ 21.0\pm  3.0$ \\
H$_2$~$J$=4 & 1035.188 & 1.068 & \phantom{1}$ 27.3\pm  3.0$ \\
H$_2$~$J$=4 & 1044.548 & 1.206 & \phantom{1}$ 30.1\pm  3.2$ \\
H$_2$~$J$=4 & 1060.588 & 1.019 & \phantom{1}$ 20.4\pm  2.3$ \\
HD~$J$=0    &  959.817 & 1.150 & \phantom{1}$ 13.4\pm  4.0$ \\
HD~$J$=0    & 1007.283 & 1.515 & \phantom{1}$ 14.8\pm  3.2$ \\
HD~$J$=0    & 1011.457 & 1.423 & \phantom{1}$ 15.9\pm  2.7$ \\
HD~$J$=0    & 1042.848 & 1.332 & \phantom{1}$ 13.6\pm  2.2$ \\
HD~$J$=0    & 1054.288 & 1.238 & \phantom{1}$ 12.8\pm  2.3$ \\
HD~$J$=0    & 1066.271 & 1.089 & \phantom{10}$  6.6\pm  2.1$ \\
\enddata
\vspace{2mm}
\tablenotetext{a}{Wavelengths and $f$-values are from the
\Owens\ and \vpfit\ atomic databases, which use molecular transition information from
\citet{Abgrall93a,Abgrall93b}.}
\end{deluxetable}

\clearpage
\begin{deluxetable}{ccc}
\tablewidth{0pc}
\tablecaption{\HH\ and HD Column Densities.
\label{tab-pg0038h2columns}}
\tablehead{ 
\colhead{\phantom{000}Species\phantom{000}} & \colhead{\phantom{000000}$J$\phantom{000000}}     &\colhead{\phantom{000}Adopted ($2\sigma$)\tablenotemark{a}}\phantom{000}}
\startdata
\HH\ & 0 &   $18.76\pm0.05$  \\
\HH\ & 1 &   $19.18\pm0.05$  \\
\HH\ & 2 &   $17.73\pm0.40$\tablenotemark{b}  \\
\HH\ & 3 &   $16.74\pm0.40$\tablenotemark{b}  \\
\HH\ & 4 &   $14.63\pm0.15$  \\
\HH\ & 5 &   $14.04\pm0.15$  \\ 
\HH\ &total & $19.33\pm 0.04$  \\ \hline
     &   &                                          \\
HD   & 0 &   $13.95\pm0.10$  \\
\enddata
\vspace{2mm}
\tablenotetext{a}{The values are composite results from the COG and PF analyses
  (using both \owens\ and \vpfit ).  Quoted errors are $2\sigma$, 
  and we assume $1\sigma$
  errors are half the $2\sigma$ uncertainties.  See text for details.}
\tablenotetext{b}{Column density errors for \HH~($J=2,3$) are uncertain due to lines being on the flat part of the curve of growth.}
\end{deluxetable}

\clearpage
\begin{deluxetable}{lrr}
\tablewidth{0pc}
\tablecaption{Atomic lines used for column density measurements
\label{tab-atomiclines}}
\tablehead{ 
\colhead{Species} & \colhead{$\lambda$ (\AA )\tablenotemark{a}}     &\colhead{$\log f\lambda$} }
\startdata
\HI	   &  919.351 & 0.043  \\
\HI	   &  920.963 & 0.168  \\
\HI	   &  923.150 & 0.312  \\
\HI	   &  926.226 & 0.471  \\
\HI	   & 1025.722 & 1.909  \\
\DI	   &  919.102 & 0.043  \\
\DI	   &  920.712 & 0.170  \\
\DI	   &  922.899 & 0.311  \\
\DI	   &  925.974 & 0.470  \\
\DI	   & 1025.443 & 1.909  \\
\CI	   &  945.191 & 2.157  \\
\CI	   & 1111.421 & 0.996  \\
\CI	   & 1122.438 & 0.820  \\
\CI	   & 1129.193 & 0.998  \\
\CI	   & 1139.793 & 1.206  \\
\CI	   & 1157.910 & 1.451  \\
\CI	   & 1158.324 & 0.810  \\
\CIstar    &  945.338 & 2.157  \\
\CII	   & 1036.337 & 2.102  \\
\CIII	   &  977.020 & 2.875  \\
\NI	   &  951.079 &-0.794  \\
\NI	   &  952.523 &-0.401  \\
\NII	   & 1083.994 & 2.072  \\
\OI	   &  974.070 &-1.807  \\
\NaI	   & 5891.583 & 3.596  \\
\NaI	   & 5897.558 & 3.296  \\
\SiII	   & 1020.699 & 1.188  \\
\PII	   & 1152.818 & 2.435  \\
\ion{S}{3} & 1012.501 & 1.634  \\
\ArI	   & 1048.220 & 2.408  \\
\FeII	   &  940.192 & 1.056  \\
\FeII	   & 1055.262 & 0.926  \\
\FeII	   & 1081.875 & 1.180  \\
\FeII      & 1096.877 & 1.545  \\
\FeII	   & 1121.975 & 1.351  \\
\FeII	   & 1125.448 & 1.093  \\
\FeII	   & 1127.098 & 0.529  \\
\FeII	   & 1133.665 & 0.833  \\
\FeII	   & 1142.366 & 0.757  \\
\FeII	   & 1143.226 & 1.182  \\
\enddata
\vspace{2mm}
\tablenotetext{a}{Wavelengths and $f$-values are from the \Owens\ and \vpfit\ atomic databases, 
which use information from
\citet{Morton91,Morton03,Howk00} and \citet{Verner94}.}
\end{deluxetable}

\clearpage
\begin{deluxetable}{llccl}
\tablewidth{0pc}
\tablecaption{Atomic Column Densities. 
\label{tab-pg0038atomiccolumns}}
\tablehead{ 
\colhead{Ion}                & \colhead{Adopted } &                             \multicolumn{2}{c}{Errors\tablenotemark{a}} &\colhead{Technique\tablenotemark{b}}\\
\colhead{}                   & \colhead{Value}                                              & \colhead{$1\sigma$} & \colhead{$2\sigma$} & \colhead{}  }\startdata
\HI\			     &  \phantom{$\ge$o}$20.41$    		                      &0.04		  &0.08 	     & PF (3)				 \\
\HI +2\HH                    &  \phantom{$\ge$o}$20.48 $                                      &0.04		  &0.07 	     & PF, COG  		 \\
\DI\			     & \phantom{$\ge$o}$15.75 $	  	                   	      &0.04		  &0.08 	     & PF (3)				 \\
\DI +HD                      &  \phantom{$\ge$o}$15.76 $                                      &0.04		  &0.08 	     & PF, COG  		 \\
\ion{C}{1}		     & \phantom{$\ge$o}$13.80 $	  	                  	      &0.04		  &0.08 	     & PF (2)			      \\
\ion{C}{1}$^*$		     & \phantom{$\ge$o}$13.25 $	                                      &0.19		  &$^{+0.32}_{-0.54}$& PF (1)				\\
\ion{C}{2}		     & $\ge$ 14.47	  					      &\nodata            &\nodata           & AOD			     \\
\ion{C}{3}		     & $\ge$ 13.95 						      &\nodata            &\nodata           & AOD			     \\
\ion{N}{1}		     & \phantom{$\ge$o}$16.39 $ 	                              &$^{+0.07}_{-0.10}$ &$^{+0.14}_{-0.18}$ & PF (3)                       \\
\ion{N}{2}		     & $\ge$ 14.12	 					      &\nodata            &\nodata           & AOD			     \\
\ion{O}{1}		     & \phantom{$\ge$o}$17.37 $	                                      &$^{+0.08}_{-0.15}$ &$^{+0.15}_{-0.25}$ & PF (3)                       \\
\ion{Na}{1}		     & \phantom{$\ge$o}$12.19 $            		              &0.03               &0.04   & AOD, PF (2)		     \\
\ion{Si}{2}		     & $\ge$ 15.17	   					      &\nodata            &\nodata            & AOD			     \\
\ion{P}{2}                   & $\ge$ 13.34						      &\nodata            &\nodata            & AOD			     \\
\ion{S}{3}                   & \phantom{$\ge$o}$13.98 $	                                      &$^{+0.03}_{-0.06}$ &$^{+0.07}_{-0.10}$ & PF (1)                       \\
\ion{Ar}{1}                  & $\ge$ 13.35    					              &\nodata            &\nodata            & AOD			     \\
\ion{Fe}{2}                  & \phantom{$\ge$o}$14.42 $                                       &$^{+0.03}_{-0.02}$ &$^{+0.06}_{-0.05}$ & PF (3),COG                   \\
\enddata 
\tablenotetext{a}{For comparison with other ISM work, we give both 1 and $2\sigma$ errors. }
\tablenotetext{b}{PF (profile fit, \owens\ or \vpfit , with number of measurements in parentheses), 
COG (curve of growth) or AOD (apparent optical depth).}
\end{deluxetable}

\clearpage
\begin{deluxetable}{lll}
\tablewidth{0pc}
\tablecaption{Unidentified lines in \FUSE\ data for \pgoo\
\label{tab-unidentifiedlines}}
\tablehead{ 
\colhead{Wavelength range (\AA )} & \colhead{Wavelength range (\AA )} & \colhead{Wavelength range (\AA )} }
\startdata
973.0 - 973.2	   &   1062.2 - 1062.4				 & 1144.7 - 1144.9    \\
976.6 - 976.8	   &   1067.0 - 1067.2				 & 1146.0 - 1146.5\tablenotemark{c}    \\
997.3 - 997.5	   &   1071.9 - 1072.1				 & 1146.55 - 1146.8   \\
999.5 - 999.8	   &   1073.65 - 1073.85			 & 1147.2 - 1147.4    \\
1014.6 - 1014.8    &   1080.35 - 1080.7\tablenotemark{a} 	 & 1147.9 - 1148.1    \\
1051.9 - 1052.1    &   1080.95 - 1081.10			 & 1173.45 - 1173.89\tablenotemark{d}  \\
1052.6 -1052.8     &   1125.15 - 1125.35			 & 1180.48 - 1180.68  \\
1061.3 - 1061.5    &   1141.1 - 1141.3\tablenotemark{b}				 &		      \\
\enddata
\vspace{2mm}
\tablenotetext{a}{2 lines}
\tablenotetext{b}{on the blue side of \ion{Fe}{2} $\lambda$1144.94}
\tablenotetext{c}{\ion{Fe}{7} $\lambda$1146.46?}
\tablenotetext{d}{3 lines}
\end{deluxetable}

\clearpage

\clearpage
\begin{deluxetable}{lccc}
\tablewidth{0pc}
\tablecaption{\DI , \HI , \OI , \NI , HD, \HH\ ratios.
\label{tab-ratios}}
\tablehead{ 
\colhead{Species} & \colhead{Ratio} & \colhead{Errors ($1\sigma$)} & \colhead{Errors ($2\sigma$)}} 
\startdata
$10^{5}$\DHrattot  & $1.91$ & $+0.26\, -0.24$ & $+0.52\, -0.42$  \\
$10^{5}$D/H & $2.19$ & $+0.30\, -0.27$ & $+0.55\, -0.50$  \\
$10^{4}$\OHtot  & $7.76$ & $+1.78\, -2.33$ & $+3.38\, -3.49$  \\
$10^{4}$O/H & $8.51$ & $+1.99\, -2.56$ & $+4.09\, -3.86$  \\
$10^{4}$\NHtot  & $0.81$ & $+0.17\, -0.18$ & $+0.36\, -0.29$  \\
$10^{4}$N/H & $0.95$ & $+0.20\, -0.20$ & $+0.43\, -0.34$  \\
$10^{2}$\DOtot  & $2.63$ & $+1.13\, -0.49$ & $+2.18\, -0.85$  \\
$10^{2}$D/O & $2.40$ & $+1.03\, -0.45$ & $+2.19\, -0.78$  \\
$10^{1}$\DNtot  & $2.34$ & $+0.66\, -0.39$ & $+1.35\, -0.72$  \\
$10^{1}$D/N & $2.29$ & $+0.65\, -0.39$ & $+1.32\, -0.71$  \\
$10^{2}$N/O & $1.12$ & $+0.52\, -0.28$ & $+1.05\, -0.47$  \\
$10^{6}$HD/\HH\ & $4.17$ & \nodata & $+0.68\, -0.49$  \\
\enddata
\vspace{2mm}
\end{deluxetable}

\clearpage

\clearpage

\clearpage

\clearpage

\begin{figure}
\begin{center}
\rotatebox{0}{
\epsscale{0.8}
\vspace*{6in}
}
\caption[Fit to \OI ]{
\Owens\ profile fit to \OI\, showing the fit before convolution with
the local LSF {\it (light blue line)} and after convolution {\it (red line)}.  
Flux is in erg~cm$^{-2}$~s$^{-1}$~\AA$^{-1}$.  See stand-alone JPEG file.
\label{fig-traceOI}}
\end{center}
\end{figure}

\clearpage

\clearpage

\clearpage

\clearpage

\clearpage

\clearpage

\clearpage


\begin{thebibliography}{}

\bibitem[Abgrall et al.(1993a)]{Abgrall93a}Abgrall, H., Roueff, E., Launay, F., Roncin, J.-Y., Subtil, J.-L. 1993,
A\&AS, 101, 273

\bibitem[Abgrall et al.(1993b)]{Abgrall93b}Abgrall, H., Roueff, E., Launay, F., Roncin, J.-Y., Subtil, J.-L. 1993,
A\&AS, 101, 323


\bibitem[Andr\'e et al.(2003)]{Andre03}Andr\'e, M. K. et al. 2003, ApJ, 591, 1000

\bibitem[Bi\'emont \& Zeippen(1992)]{Biemont92}Bi\'emont, E., \& Zeippen, C. S. 1992, A\&A, 265, 850

\bibitem[Black \& Dalgarno(1973)]{Black73}Black, J. H. \& Dalgarno, A. 1973, ApJ, 184, L101

\bibitem[Burles(2000)]{Burles00}Burles, S. 2000, Nucl. Phys. A, 663, 861

\bibitem[Burles et al.(2001)]{Burles01}Burles, S., Nollett, K. M., \& Turner, M. S.  2001, ApJ, 552, L1

\bibitem[Carswell et al.(1996)]{Carswell96}Carswell, R. F., Webb, J. K., Lanzetta, K. M., Baldwin, J. A., 
Cooke, A. J., Williger, G. M., Rauch, M., Irwin, M. J., Robertson, J. G., Shaver, P. A. 1996, MNRAS, 278, 506

\bibitem[Cartledge et al.(2004)]{Cartledge04}Cartledge, S. I. B., Lauroesch, J. T., Meyer, D. M., Sofia, U. J. 2004, ApJ, in press


\bibitem[Cha et al.(2000)]{Cha00}Cha, A. N. S., Sahu, M. S., Moos, H. W., Blaauw, A. 2000, ApJS, 129, 281

\bibitem[Chiappini, Renda, \& Matteucci(2002)]{Chiappini02}Chiappini, C., Renda, A., \& 
Matteucci, F. 2002, A\&A, 395, 789

\bibitem[Crighton et al.(2003)]{Crighton03}Crighton, N. H. M., Webb, J. K., Carswell, R. F., Lanzetta, K. M. 2003,
MNRAS, 345, 243

\bibitem[Cyburt et al.(2003)]{Cyburt03}Cyburt, R., H., Fields, B. D., \& Olive, K. A. 2003, Physics Letters B, 567, 227

\bibitem[deAvillez \& MacLow(2002)]{deAvillez02}deAvillez, M. A., \& MacLow, M.-M. 2002, ApJ, 581, 1047

\bibitem[Draine(2004)]{Draine04}Draine, B. T. 2004, to appear in ``Carnegie Observatories Astrophysics Series, Vol. 4: 
Origin and Distribution of the Elements", ed. A. McWilliam and M. Rauch (Cambridge: Cambridge Univ. Press),
astro-ph/0312592



\bibitem[Dreizler et al.(1997)]{Dreizler97}Dreizler, S., Werner,
K., Heber, U., Reid, N.,  Hagen, H.\ 1997, in The Third Conference on Faint
Blue Stars, ed. A. G. D. Philip, J. Liebert, R. Saffer, D. S. Hayes (L. Davis Press), 303

\bibitem[Dreizler \& Werner(1996)]{Dreizler96}Dreizler, S., \& Werner, K. 1996, A\&A, 314, 217

\bibitem[Ferlet et al.(2000)]{Ferlet00}Ferlet, R. et al. 2000, ApJ, 538, 69

\bibitem[Ferlet, Vidal-Madjar \& Gry(1985)]{Ferlet85}Ferlet, R., Vidal-Madjar, A., \& Gry, C. 1985,
ApJ, 298, 838

\bibitem[Fitzpatrick \& Spitzer(1994)]{Fitzpatrick94}Fitzpatrick, E. L. \& Spitzer, L., Jr. 1994, ApJ, 427, 232

\bibitem[Friedman et al.(2002)]{Friedman02}Friedman, S. D., et al. 2002, ApJS, 140, 37

\bibitem[Green et al.(1986)]{Green86}Green, R. F., Schmidt, M., \& Liebert J. 1986, ApJS, 61, 305

\bibitem[Haffner et al.(2003)]{Haffner03}Haffner, L. M., Reynolds, R. J., Tufte, S. L., Madsen, G. J.,
Jaehnig, K. P., Percival, J. W. 2003, ApJS, 149, 405

\bibitem[Hartmann \& Burton(1997)]{Hartmann97}Hartmann, L. \& Burton, W. B. 1997,
Atlas of Galactic Neutral Hydrogen. Cambridge Univ. Press, ISBN 0521471117

\bibitem[H\'ebrard et al.(2002)]{Hebrard02}H\'ebrard, G. et al., ApJS, 140, 103

\bibitem[H\'ebrard \& Moos(2003)]{Hebrard03}H\'ebrard, G. \& Moos, H. W. 2003, ApJ, 599, 297

\bibitem[H\'ebrard et al.(2004)]{Hebrard04}H\'ebrard, G., Chayer, P., Dupuis, J.,
Moos, H. W., Sonnentrucker, P., Tripp, T. M., Williger, G. M. 2004, 
Astrophysics in the Far UV: Five Years of Discovery with FUSE,
ed. Sonneborn, G., Moos, H. W. \& Andersson, B.-G., ASP Conf. Series, in prep.

\bibitem[Hoopes et al.(2003)]{Hoopes03}Hoopes, C. G., Sembach, K. R., H\'ebrard, G., Moos, H. W., \& Knauth, D. C.
2003, ApJ, 586, 1094

\bibitem[Howk et al.(2000)]{Howk00}Howk, J. C., Sembach, K. R., Roth, K. C., Kruk, J. W. 2000, ApJ, 544, 867

\bibitem[Hubeny \& Lanz(1995)]{Hubeny95}Hubeny, I., \& Lanz, T. 1995, ApJ, 439, 875 

\bibitem[Jenkins(1986)]{Jenkins86}Jenkins, E. B. 1986, ApJ, 304, 739

\bibitem[Jenkins, Savage, \& Spitzer(1986)]{Jenkins86b}Jenkins, E.B., Savage, B.D., \&
Spitzer, L., Jr. 1986, ApJ, 301, 355

\bibitem[Jenkins et al.(1999)]{Jenkins99}Jenkins, E. B., Tripp, T. M., Wo\.{z}niak, P. R., Sofia, U. J., \& Sonneborn,
G. 1999, ApJ, 520, 182

\bibitem[Jenkins et al.(2000)]{Jenkins00}Jenkins, E. B., Wo\.{z}niak, P. R., Sofia, U. J., Sonneborn,
G., Tripp, T. M. 2000, ApJ, 538, 275

\bibitem[Kirkman et al.(2003)]{Kirkman03}Kirkman, D., Tytler, D., Suzuki, N., O'Meara, J. M., \& Lubin, D. 2003, ApJS, 149, 1

\bibitem[Kohoutek(1997)]{Kohoutek97}Kohoutek, L. 1997, Astron. Nachr., 318, 35

\bibitem[Kruk et al.(2002)]{Kruk02}Kruk, J. W., et al. 2002, ApJS, 140, 19

\bibitem[Lacour et al.(2004)]{Lacour04}Lacour, S., Andr\'e, M. K., Sonnentrucker,
P., LePetit, F., Welty, D. E., Desert, J.-M., Ferlet, R., Roueff, E.,
York, D. G. 2005, A\&A, 430, 967

\bibitem[Laurent et al.(1979)]{Laurent79}Laurent, C., Vidal-Madjar, A., \& York, D. G. 1979, ApJ, 229, 923

\bibitem[Lemoine et al.(1999)]{Lemoine99}Lemoine, M. et al. 1999, NewA, 4, 231

\bibitem[Lemoine et al.(2002)]{Lemoine02}Lemoine, M. et al. 2002, ApJS, 140, 67

\bibitem[Levshakov et al.(2002)]{Levshakov02}Levshakov, S. A., Dessanges-Zavadsky, M., D'Odorico, S., \& Molaro, P. 2002, ApJ, 565, 696

\bibitem[Linsky(1998)]{Linsky98}Linsky, J. L. 1998, Space Sci. Rev., 84, 285


\bibitem[Lipshtat et al.(2004)]{Lipshtat04}Lipshtat, A., Ofer, B., \& Herbst, E. 2004, MNRAS, 348, 1055

\bibitem[Liszt(2003)]{Liszt03}Liszt, H. 2003, A\&A, 398, 621

\bibitem[Messenger(2000)]{Messenger00}Messenger, S. 2000, Nature, 404, 968

\bibitem[Meyer, Cardelli, \& Sofia(1997)]{Meyer97}Meyer, D. M., Cardelli, J. A., \& Sofia, U. J. 1997, ApJ, 490, L103

\bibitem[Meyer, Jura, \& Cardelli(1998)]{Meyer98}Meyer, D. M., Jura, M., \& Cardelli, J. A. 1998, ApJ, 493, 222

\bibitem[Morton(1991)]{Morton91}Morton, D. C., 1991, ApJS, 77, 119

\bibitem[Morton(2003)]{Morton03}Morton, D. C., 2003, ApJS, 149, 205

\bibitem[Moos et al.(2000)]{Moos00}Moos, H. W. et al. 2002, ApJ, 538, L1

\bibitem[Moos et al.(2002)]{Moos02}Moos, H. W. et al. 2002, ApJS, 140, 3

\bibitem[Oliveira et al.(2003)]{Oliveira03}Oliveira, C. M., H\'ebrard, G., Howk, J. K., Kruk, J. W., Chayer, P., \& Moos, H. W. 2003, ApJ, 587, 235

\bibitem[O'Meara et al.(2001)]{Omeara01}O'Meara, J. M., Tytler, D., Kirkman, D., Suzuki, N., Prochaska, J. X., Lubin, D., \& Wolfe, A. M. 2001, ApJ, 552, 718

\bibitem[Peeters et al.(2004)]{Peeters04}Peeters, E., Allamandola, L. J., Bauschlicher, C. W., Jr., Hudgins, D. M.,
Sandford, S. A., Tielens, A. G. G. M. 2004, ApJ, 604, 252

\bibitem[Pettini \& Bowen(2001)]{Pettini01}Pettini, M., \& Bowen, D. V. 2001, ApJ, 560, 41

\bibitem[Rachford et al.(2002)]{Rachford02}Rachford, B. L. et al. 2002, ApJ, 577, 221

\bibitem[Rachford et al.(2001)]{Rachford01}Rachford, B. L. et al. 2001, ApJ, 555, 839


\bibitem[Rauch, Kerber, \& Pauli(2004)]{Rauch04}Rauch, T., Kerber, F., \& Pauli, E.-M.\ 2004, A\&A, 417, 647 


\bibitem[Romano et al.(2003)]{Romano03}Romano, D., Tosi, M., Matteucci, F., \& Chiappini, C. 2003, MNRAS, 346, 295

\bibitem[Sahnow et al.(2000)]{Sahnow00}Sahnow, D. J., et al. 2000, ApJ, 538, L7

\bibitem[Savage et al.(1977)]{Savage77}Savage, B. D., Drake, J. F., Budich, W., \&
Bohlin, R. C. 1977, ApJ, 216, 291

\bibitem[Savage \& Mathis(1979)]{Savage79}Savage, B. D. \& Mathis, J. S. 1979,
ARAA, 17, 73

\bibitem[Savage \& Sembach(1991)]{Savage91}Savage, B. D., \& Sembach, K. R. 1991, ApJ, 379, 245

\bibitem[Schlegel et al.(1998)]{Schlegel98}Schlegel, D. J., Findbeiner, D. P., \& Davis, M. 1998, ApJ, 500, 525

\bibitem[Schramm \& Turner(1998)]{Schramm98}Schramm, D. N., \& Turner, M. S. 1998, Rev. Mod. Phys., 70, 303

\bibitem[Sembach \& Savage(1992)]{Sembach92}Sembach, K. R., \& Savage, B. D. 1992,  ApJS, 83, 147

\bibitem[Sembach et al.(2004)]{Sembach04}Sembach, K. R., et al. 2004, ApJS, 150, 387

\bibitem[Sfeir et al.(1999)]{Sfeir99}Sfeir, D. M., Lallement, R., Crifo, F., \& Welsh, B. Y. 1999, A\&A, 346, 785

\bibitem[Snow et al.(2000)]{Snow00}Snow, T. P. et al. 2000, ApJ, 538, 65

\bibitem[Snowden et al.(1995)]{Snowden95}Snowden, S. L. et al. 1995, ApJ, 454, 643

\bibitem[Sonneborn et al.(2000)]{Sonneborn00}Sonneborn, G., Tripp, T. M., Ferlet, R., Jenkins, E. B.,
Sofia, U. J., Vidal-Madjar, A., Wozniak, P. R. 2000, ApJ, 545, 277

\bibitem[Spitzer(1985)]{Spitzer85}Spitzer, L., Jr. 1985, ApJ, 290, L21

\bibitem[Spitzer, Cochran, \& Hirschfeld(1974)]{Spitzer74}Spitzer, L., Jr., Cochran, W. D., \& 
Hirschfeld, A. 1974, ApSS, 28, 373

\bibitem[Steigman(2003)]{Steigman03}Steigman, G. 2003, ApJ, 586, 1120

\bibitem[Stephenson \& Green(2002)]{Stephenson02}Stephenson, F. R. \& Green, D. A. 2002, ``Historical Supernovae and
their Remnants" (Oxford: Clarendon Press),  ISBN 0198507666

\bibitem[Verner, Barthel, \& Tytler(1994)]{Verner94}Verner, D. A., Barthel, P. D., \& Tytler, D. 1994, A\&AS,
108, 287

\bibitem[Vogt et al.(1994)]{Vogt94} Vogt, S.~S., et al.\ 1994,  \procspie , 2198, 362 

\bibitem[Webb(1987)]{Webb87}Webb, J. K. 1987, PhD thesis, Cambridge University

\bibitem[Webb(1991)]{Webb91}Webb, J. K., Carswell, R. F., Irwin, M. J., Penston, M. V. 1991, MNRAS, 250, 657

\bibitem[Welty et al.(1999)]{Welty99}Welty, D. E., Hobbs, L. M., Lauroesch, J. T.,
Morton, D. C., Spitzer, L., Jr., York, D. G. 1999, ApJS, 124, 465

\bibitem[Werner et al.(1997)]{Werner97}Werner, K., Bagschik, K., Rauch, T.,  Napiwotzki, R.\ 1997, A\&A, 327, 
721 

\bibitem[Wesemael et al.(1993)]{Wesemael93}Wesemael, F., Greenstein, J. L., Liebert, J., Lamontagne, R., 
Fontaine, G., Bergeron, P., \& Glaspey, J. W. 1993, PASP, 105, 761

\bibitem[Wiese, Fuhr \& Deters(1996)]{Wiese96}Wiese, W. L., Fuhr, J. R., \& 
Deters, T. M. 1996, J. Phys. Chem. Ref. Data Monogr. 7

\bibitem[Williams et al.(2001)]{Williams01}Williams, T., McGraw, J.T., Mason, P.A., \& Grashius, R. 2001, PASP, 113, 490

\bibitem[Wood et al.(2002)]{Wood02}Wood, B. E., Linsky, J. L., H\'ebrard, G., Vidal-Madjar, A., Lemoine, M., 
Moos, H. W., Sembach, K. R., \& Jenkins, E. B. 2002, ApJS, 140, 91

\bibitem[Wood et al.(2004)]{Wood04}Wood, B. E., Linsky, J. L., H\'ebrard, G., Williger, G. M., Moos,
H. W., \& Blair, W. P. 2004, ApJ, 609, 838

\bibitem[York \& Rogerson(1976)]{York76}York, D. G., \& Rogerson, J. B. 1976, ApJ, 203, 378

\bibitem[York(1983)]{York83}York, D. G. 1983, ApJ, 264, 172

\bibitem[Zuckerman \& Reid(1998)]{Zuckerman98} Zuckerman, B.~\& Reid, I.~N.\ 1998, \apjl , 505, L143 

\end{thebibliography}
\end{document}